\documentclass[apj,numberedappendix]{emulateapj}
\usepackage{lscape}

\newcommand{\eg}{e.g.}

\newcommand{\fig}[1]{Fig.~\ref{#1}}

\newcommand{\ie}{i.e.}

\newcommand{\ctwo}{\ion{C}{2}}
\newcommand{\cthree}{\ion{C}{3}}

\newcommand{\halpha}{H$\alpha$}

\newcommand{\hone}{\ion{H}{1}}
\newcommand{\htwo}{H$_2$}

\newcommand{\oone}{\ion{O}{1}}
\newcommand{\osix}{\ion{O}{6}}

\newcommand{\ebv}{$E($\bv)}

\newcommand{\flux}{erg cm$^{-2}$ s$^{-1}$ \AA$^{-1}$}
\newcommand{\kms}{km s$^{-1}$}

\newcommand{\sig}{$\sigma$}
\newcommand{\specfit}{{\small SPECFIT}}

\newcommand{\fuse}{{\it FUSE}}

\newcommand{\iras}{{\it IRAS}}
\newcommand{\iue}{{\it IUE}}
\newcommand{\rosat}{{\it ROSAT}}
\newcommand{\spear}{{\it SPEAR}}

\journalinfo{To appear in {\it The Astrophysical Journal,} Vol. 647, 10 August 2006}
\submitted{To appear in {\rm The Astrophysical Journal,} Vol. 647, 10 August 2006}

\shorttitle{\osix\ Emission in the ISM}
\shortauthors{Dixon et al.}

\begin{document}

\title{An Extended {\em FUSE}\/ Survey of Diffuse O VI Emission in the Interstellar Medium$^1$}

\footnotetext[1]{Based on observations made with the NASA-CNES-CSA {\it
Far Ultraviolet Spectroscopic Explorer.  FUSE}\/ is operated for NASA by
the Johns Hopkins University under NASA contract NAS5-32985.}

\author{W.\ Van Dyke Dixon and Ravi Sankrit}
\affil{The Johns Hopkins University, Department of Physics and Astronomy,
3400 North Charles Street, Baltimore, MD 21218}
\email{wvd@pha.jhu.edu}

\and

\author{Birgit Otte}
\affil{University of Michigan, Astronomy Department,
500 Church Street, Ann Arbor, MI 48109}

\begin{abstract}
We present a survey of diffuse \osix\ emission in the interstellar
medium obtained with the {\it Far Ultraviolet Spectroscopic Explorer
(FUSE).}\/  Spanning 5.5 years of \fuse\/ observations, from launch
through 2004 December, our data set consists of 2925 exposures
along 183 sight lines, including all of those with previously-published
\osix\ detections.  The data were processed using an implementation
of CalFUSE v3.1 modified to optimize the signal-to-noise ratio and
velocity scale of spectra from an aperture-filling source.  Of our
183 sight lines, 73 show \osix\ $\lambda 1032$ emission, 29 at $> 3 \sigma$
significance.  Six of the $3 \sigma$ features have velocities $|v_{\rm LSR}| > 120$
\kms, while the others have $|v_{\rm LSR}| < 50$ \kms.  Measured
intensities range from 1800 to 9100 LU, with a median of 3300 LU.
Combining our results with published \osix\ absorption data, we find that 
an \osix -bearing interface in the local ISM yields an electron density 
$n_{\rm e}$ = 0.2--0.3 cm$^{-3}$ and a path length of 0.1 pc, while
\osix -emitting regions associated with high-velocity clouds in the Galactic halo
have densities an order of magnitude lower 
and path lengths two orders of magnitude longer.  Though the \osix\ intensities along these
sight lines are similar, the emission is produced by gas with very different properties.

\end{abstract}


\section{Introduction}

For gas in collisional ionization equilibrium, emission via the
1031.93 and 1037.62~\AA\ resonance lines of the lithium-like \osix\
ion is the dominant cooling mechanism at temperatures of (1--5$)
\times 10^5$ K \citep{Sutherland:Dopita93}.  Gas cools rapidly at
these temperatures, so \osix\ in the interstellar medium (ISM)
traces regions in transition: hot gas cooling through temperatures
of a few times $10^5$ K or interfaces between cool or warm gas ($T
= 10^2$--$10^4$ K) and hot gas ($T = 10^6$ K) where $10^5$ K gas
can form \citep{Savage:95}.  

Absorption-line studies have begun to reveal the distribution of
\osix -bearing gas in the Galaxy. 
(For an excellent review, see \citealt{Savage:06}.)
\ion{O}{6} absorption is detected
in the spectra of UV-bright stars, QSOs, and AGNs
\citep[\eg,][]{Wakker:03}. Measurements along sight lines
through the Galactic halo indicate that the \osix -bearing gas is
roughly co-spatial with the thick disk, having a scale height of
about 2.3 kpc \citep{Savage:03}.  Within the thick disk, the
distribution of \ion{O}{6} is patchy and varies on small angular
scales \citep[$0\fdg05-5\fdg0$ toward the Magellanic Clouds;][]
{Howk:02}. Measurements towards stars in the disk indicate that the
\osix -bearing gas is extremely clumpy and cannot exist
in uniform clouds \citep{Bowen:06}. These observations are consistent
with the \osix\ being formed in interfaces \citep{Savage:06}.

Emission-line studies provide additional insight into the properties 
of transition-temperature gas.
While absorption-line studies reveal the velocity distribution of
\osix -bearing gas along a line of sight, they are limited to sight
lines with bright background sources.  Emission-line observations
can probe (and eventually map) the entire sky, but with
lower spectral and spatial resolution.  Particularly useful are
measurements of \osix\ absorption and emission along a single line
of sight. The absorption is proportional to the density of the gas,
while the emission is proportional to the square of the density.
If the same gas is responsible for both absorption and emission,
these measurements can be combined to derive the electron density in
the plasma \citep{Shull:Slavin:94}.  Assuming a gas temperature and
oxygen abundance, one can derive the \osix\ density and path length
through the emitting gas.  

Until recently, convincing detections of \osix\ emission from the
diffuse interstellar medium were limited to fewer than a dozen sight
lines probed with the {\it Far Ultraviolet Spectroscopic Explorer}
({\it FUSE;}\/ Table \ref{tab_published}).  Reported intensities
range from 1.6 to $3.3 \times 10^3$ LU.  (One photon cm$^{-2}$
s$^{-1}$ sr$^{-1}$ or line unit corresponds to $1.9 \times 10^{-11}$
erg cm$^{-2}$ s$^{-1}$ sr$^{-1}$ at 1032 \AA.) \citet{Korpela:06}
present a deep far-ultraviolet emission spectrum from a region of
15\degr\ radius centered on the north ecliptic pole obtained with
the {\it Spectroscopy of Plasma Evolution from Astrophysical Radiation
(SPEAR)}\/ instrument.  They report a combined \osix\ $\lambda
\lambda 1032, 1038$ intensity of 5724 LU, slightly higher than the
initial \fuse\/ results.  Using archival data from the first four
years of the \fuse\/ mission, \citet{Otte:06} find measurable \osix\
$\lambda 1032$ emission along 23 of 112 sight lines and conclude
that their data are consistent with the picture derived from \osix\
absorption surveys: high-latitude sight lines probe \osix -emitting
gas in a clumpy, thick disk, while low-latitude sight lines sample
mixing layers and interfaces in the thin disk of the Galaxy.
Unfortunately, the small size and low signal-to-noise (S/N) ratio
of their sample (only 11 of their \osix\ features are 3\sig\
detections) limit their ability to constrain the properties of the
emitting gas.

To better constrain the physical properties of \osix -bearing gas
in the ISM, we have conducted an extended \fuse\/ survey of diffuse
\osix\ emission in the interstellar medium, using all \fuse\/ data
obtained through 2004 December and the latest version of the CalFUSE
calibration pipeline.  The results are presented in this paper,
which is organized as follows:  In \S \ref{SEC_OBSERVATIONS}, we
discuss the \fuse\/ instrument and our selection of survey sight
lines from the \fuse\/ archive.  In \S \ref{SEC_DATA}, we describe
our data-reduction techniques, with emphasis on our modifications
to the standard CalFUSE pipeline.  \S \ref{SEC_MEASUREMENTS} describes
our method for identifying \osix\ emission features and measuring
their parameters.  We discuss our results in \S \ref{SEC_RESULTS}.
In \S \ref{SEC_DISCUSSION}, we combine \osix\ emission and absorption
data to derive the properties of \osix -bearing gas in the Galactic
disk and thick disk/halo.  Results are summarized in \S \ref{SEC_SUMMARY}.
An appendix discusses the origin of the emission seen toward sight
lines P12011 and B12901.  Unless otherwise noted, all wavelengths
in this paper are heliocentric, and all velocities are quoted
relative to the local standard of rest (LSR).  We use the kinematical
LSR, in which the standard solar motion is 20 \kms\ towards $\alpha$
= 18h, $\delta$ = +30\degr\ (1900).

\begin{deluxetable}{lcccccc}
\tablewidth{0pt}
\tablecaption{\label{tab_published} Published \fuse\/ \ion{O}{6} $\lambda1032$ Measurements and Upper Limits}
\tablehead{\colhead{Sight} & \colhead{$l$} & \colhead{$b$} &
\colhead{$I_{1032}$\tablenotemark{a}} & \colhead{FWHM} & \colhead{$v_{\rm LSR}$}\\
\colhead{Line} & \colhead{(deg.)} & \colhead{(deg.)} & \colhead{(10$^3$\,LU)} &
\colhead{(km\,s$^{-1}$)} & \colhead{(km\,s$^{-1}$)} & \colhead{Ref.}}
\startdata
A11701  &  \phn57.6   & $+ 88.0  $ & $   2.0\pm  0.6$ & $ 23  \pm 55  $ & $ -17  \pm 9  $ & 1 \\
S40548  &  \phn95.4   & $+ 36.1  $ & $   1.6\pm  0.3$ & $ 75  \pm  3  $ & $ -50  \pm30  $ &  2 \\
S40561  &  \phn99.3   & $+ 43.3  $ &$\leq  1.6        $ & \nodata       & \nodata       &  2 \\
P11003  & 113.0   & $+ 70.7  $ & $   2.6\pm  0.4$ & $ 75          $ & $  10         $ &  3 \\
B00303  & 156.3   & $+ 57.8  $ & $   3.3\pm  1.1$ & $210          $ & $ -51  \pm30  $ &  4 \\
B00302  & 162.7   & $+ 57.0  $ & $   2.5\pm  0.7$ & $150          $ &$ -16  \pm22  $ &  4 \\
B12901  & 278.6   & $ -45.3  $ &$<   0.5        $ & \nodata         & \nodata         &  5 \\
A11703  & 284.0   & $+ 74.5  $ & $   2.9\pm  0.7$ & $ < 80         $ & $  84  \pm15  $ &  1 \\
I20509  & 315.0   & $ -41.3  $ & $   2.9\pm  0.3$ & $160          $ & $  64         $ &  6 \\
\enddata
\tablenotetext{a}{1\,LU = 1~photon~s$^{-1}$\,cm$^{-2}$\,sr$^{-1}$.}
\tablerefs{(1) \citealt{DixonOVI:01}; (2) \citealt{Otte:03}; (3) \citealt{Shelton:02}; 
(4) \citealt{Welsh:02}; (5) \citealt{Shelton:03}; (6) \citealt{Shelton:01}}
\end{deluxetable}

\section{Observations\label{sec_observations}}

The \fuse\/ instrument consists of four independent optical paths.
Two employ LiF optical coatings and are sensitive to wavelengths
from 990 to 1187 \AA, and two use SiC coatings, which provide
reflectivity to the Lyman limit.  The four channels overlap between
990 and 1070~\AA.  Each channel possesses three apertures that
simultaneously sample different parts of the sky.  The low-resolution
(LWRS) aperture is $30\arcsec\times30\arcsec$ in size.  The
medium-resolution (MDRS) aperture spans $4\arcsec\times20\arcsec$
and lies about $3\farcm5$ from the LWRS aperture.  The high-resolution
(HIRS) aperture lies midway between the MDRS and LWRS apertures and
samples an area of $1\farcs25\times20\arcsec$.  A fourth location,
the reference point (RFPT), is offset from the HIRS aperture by
60$\arcsec$.  When a star is placed at the reference point, all
three apertures sample the background sky.  For a complete description
of \fuse, see \citet{Moos:00} and \citet{Sahnow:00}.

The data sets used in our survey fall into three categories.  First
are the S405, S505, and Z907 programs.  S405 and S505 represent
background observations, some near \fuse\/ targets (with the target
at the RFPT), others with the LWRS aperture centered on the orbit
pole.  The Z907 program consists of extragalactic targets intended
as background sources for absorption-line studies.  When these
targets are sufficiently faint, we include them in our sample.
Second, we searched the \anchor{http://archive.stsci.edu}{Multimission
Archive at Space Telescope (MAST)} for MDRS and HIRS observations
of point sources obtained in time-tag mode, as their LWRS apertures
should sample only background radiation.  Third, we include all
\fuse\/ sight lines with previously-published detections of diffuse
\osix\ emission (Table \ref{tab_published}).

We exclude from
our sample all sight lines probing known supernova remnants or
planetary nebulae, since these structures do not represent the
diffuse interstellar medium.  Sight lines that probe the Magellanic
Clouds (Sankrit et al., in preparation), the Coalsack Nebula
\citep{Andersson:04}, and the emission nebula around KPD 0005+5106
\citep*{Otte:04} are presented elsewhere and are not included here.
This sample is a superset of that presented by \citet*{Otte:06},
which includes only data obtained though 2003 July.

We use only data from the LiF 1A channel in our analysis.  Because
its sensitivity at 1032 \AA\ is more than twice that of any other
channel, including data from other channels would reduce the
signal-to-noise ratio of the resulting spectrum.  We use only data
obtained in time-tag mode, which preserves arrival time and
pulse-height information for each photon event.  This is the default
observing mode for faint targets.  Previous observers
\citep*[\eg,][]{Shelton:01,Shelton:02,Otte:03} have detected faint
(presumably geocoronal) emission on the blue wing of the \osix\
$\lambda 1032$ line in \fuse\/ spectra obtained during orbital day,
so we use only data obtained during orbital night.  Individual exposures
with less than 10 s of night exposure time are excluded from the survey,
as are combined data sets with less than 1500 ks of total night exposure time or 
with significant continuum flux.  Finally, we use only data obtained
from launch (1999 June) through 2004 December.

Table \ref{tab_data} lists the 183 sight lines in our survey and
the \fuse\/ observations contributing to each.  Each observation consists
of many exposures; our sample contains 2925 exposures from 375 observations.  
Note that we combine
data from multiple observations --- and sometimes multiple science
programs --- that sample the same or nearby lines of sight.  
The orientation of the \fuse\/ spacecraft is specified by four
quantities: the right ascension and declination of the target, the
aperture in which the target is centered, and the astronomical roll
angle (east of north) of the spacecraft about the target (and thus
the center of the target aperture).  The roll angle is constrained
by operational requirements and varies throughout the year.  Because
we combine data sets without regard to the original target aperture
(HIRS, MDRS, LWRS, or RFPT) or roll angle, a sight line listed in
Table \ref{tab_data} may represent data obtained from regions of
the sky separated by up to 7 arcmin (twice the distance between the
LWRS and MDRS apertures).  

\begin{figure}
\epsscale{1.25}
\plotone{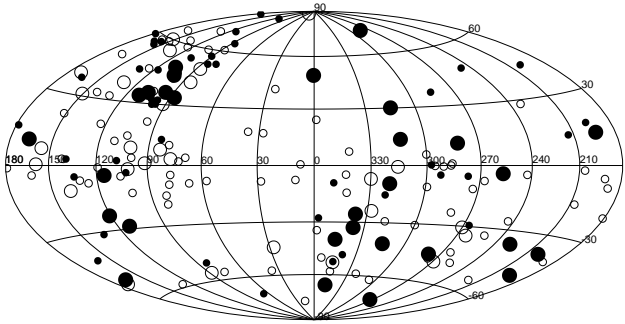}
\caption{Distribution of survey sight lines on the sky. Large solid circles represent 3\sig\ detections.  Small solid circles represent 2\sig\ detections.  Large open circles represent non-detections with upper limits less than 2000 LU.  Small open circles represent non-detections with higher upper limits.  Galactic coordinates are used in a Hammer-Aitoff projection.}
\label{fig_map}
\end{figure}

The distribution of survey sight lines on the sky is presented in
\fig{fig_map}.  The survey sample of \citet{Otte:06} is concentrated
in two quadrants, the northern sky with $0\degr < l < 180\degr$ and
the southern sky with $180\degr < l < 360\degr$, due to observational
constraints that reduced target availability in the orbit plane,
which stretches across the Galactic center.  Improvements in pointing
control implemented later in the mission have partially filled in
the other two quadrants, but the distribution remains skewed.  Note
especially the high concentration of targets in the region with
$90\degr < l < 180\degr$ and $b > 30$\degr.  This 1/16 of the sky
contains more than twice as many targets as any other region of
equal area.

\section{Data Reduction\label{sec_data}}

The data are reduced using an implementation of CalFUSE v3.1, the
latest version of the \fuse\/ data-reduction software package
(Dixon et al., in preparation), optimized for a faint, diffuse
source.  Specifically, the pipeline is instructed to reject data
obtained during orbital day.  The first three modules of the pipeline
(cf\_ttag\_init, cf\_convert\_to\_farf, and cf\_screen\_photons)
are run as usual, then the file-header keywords SRC\_TYPE and
APERTURE are set to EE and LWRS (to indicate an extended, emission-line
source in the low-resolution aperture), respectively.  Prematurely
modifying these keywords confuses the screening routines, which can
misinterpret a star drifting into the HIRS or MDRS aperture as a
detector burst.  The rest of the pipeline is run as usual, but with
background subtraction turned off.  Note that CalFUSE does not
perform jitter correction, astigmatism correction, or optimal
extraction on extended sources.

Apart from the above exceptions, we accept all of the default
parameters defined by the pipeline.  In particular, we accept all
photon events with pulse heights in the range 2--25.  To minimize
the detector background, previous observers have imposed tighter
pulse-height constraints, but we find that, because the mean pulse
height of real photon events varies with time, reducing the range
of allowed pulse heights can result in the rejection of real photon
events for some time periods.  We also accept the default spectral
binning of 0.013 \AA, or approximately two detector pixels.  
The pipeline operates on one
exposure at a time, producing a flux- and wavelength-calibrated
spectrum from each.  Error bars are computed
assuming Gaussian ($\sqrt{N}$) statistics.

The \fuse\/ wavelength scale is nominally heliocentric, but nonlinearities in early versions of the CalFUSE wavelength solution forced previous observers to derive their wavelength scales from measurements of nearby \ion{O}{1} airglow lines.  
The wavelength solution employed by CalFUSE v3.1 is far more accurate, but uncertainties in its zero point must be corrected by hand.
We fit a synthetic emission feature (described below) to the Lyman $\beta$ airglow line of each extracted spectrum and compute the shift in pixels necessary to place the line at zero velocity in a geocentric reference frame.  
(Actually, we shift the Lyman $\beta$ line to $v_{\rm geo} = +0.5$ \kms, which places the \oone\ $\lambda 1027$ line at rest.  The fainter \oone\ line is presumed to be less subject to detector effects that might skew its measured centroid.)
Zero-point wavelength shifts are not random, but exhibit a periodicity on timescales of several hours.  
We take advantage of this fact for spectra with weak Lyman $\beta$ features: if the line contains fewer than 200 raw counts, we do not attempt a fit, but interpolate a shift from the values computed for the other exposures.
The spectra are combined using the program cf\_combine, which shifts and sums the individual extracted spectral files.

\section{Measurements\label{sec_measurements}}

\subsection{Emission Line Profile}

\citet{Shelton:01} adopt the observed profile of the \oone\ $\lambda 1039$ airglow feature as the shape of an aperture-filling diffuse emission feature.  We prefer to use a synthetic line profile, so fit this feature with a model emission line consisting of a top-hat function convolved with a Gaussian.  The widths of both components are free parameters in the fit, as are the line intensity and centroid.  Fits to 131 spectra with at least 500 counts in the line yield a best-fit width of $106.1 \pm 3.4$ \kms\ for the top hat and a FWHM = $25.4 \pm 5.4$ \kms\ for the Gaussian (where the error bars represent the standard deviation about the mean).  Fits to the \oone\ $\lambda 1027$ airglow line yield similar results.  The top hat represents the projection of the LWRS aperture onto the detector, while the Gaussian represents the finite resolution of the instrument.  For point sources, the \fuse\/ resolution is between 15 and 20 \kms, so a value of 25 \kms\ for a diffuse source is reasonable.  Note that 106 \kms\ corresponds to 28 spectral pixels.

Several previous observers have followed \citet{DixonOVI:01} in adopting an unconvolved top hat for the shape of a diffuse emission feature, convolving it with a Gaussian to match the observed \osix\ line profile, and reporting the FWHM of the Gaussian as the intrinsic width of the emission feature.  By ignoring the instrumental contribution to the smoothing of the top-hat function, this technique over-estimates the intrinsic width of the emission profile.  Fortunately, the error is small: 10\% for a best-fit FWHM of 50 \kms, 1\% for 150 \kms.

To construct our model line profiles, we convolve the 106 \kms\ top-hat function with a single Gaussian, representing the combined effects of the intrinsic emission profile and the instrumental line-spread function.  To optimize the model resolution, we employ a grid of 0.013 \AA\ pixels and, rather than binning the model to match the data, smooth it by convolving with a second top-hat function either 8 or 14 pixels wide.  Each curve is normalized so that the best-fit scale factor reported by our fitting routine equals the line intensity in LU.  We generate a series of curves with Gaussian FWHM values from 1 to 1000 \kms.

\subsection{Detection of \osix\ Emission}

Following \citet{Martin:Bowyer:90}, we use an automated routine to search each composite spectrum for a statistically-significant emission feature near 1032 \AA.  The calculation is performed using the WEIGHTS array, which is effectively raw counts for low count-rate data.  (Dixon et al., in preparation, discuss the format of \fuse\/ calibrated spectral files.)  The mean value of the WEIGHTS array in the regions 1029--1030 and 1033.5--1036 \AA\ is adopted as the local continuum.  At each pixel between 1030 and 1034 \AA, we bin the data by the width of an emission feature and determine the counts in excess of the mean.  The significance of this excess is computed assuming Gaussian statistics.  We repeat this process for bin widths from 25 to 30 pixels, or $\sim$ 94 to 113 \kms.  
From all combinations of central wavelength and line width, we select the most significant feature.  If its significance is greater than $3 \sigma$, we record it as a detection.  Table \ref{tab_detections} lists our detection sight lines, Table \ref{tab_limits} our non-detection sight lines.  Our algorithm is designed to detect emission features that are $\sim$ 106 \kms\ in width; given our S/N, much narrower features are likely to be noise spikes, while much broader features are difficult to distinguish from the background.

Further analysis is performed on the flux-calibrated spectra, which are first binned to improve their signal-to-noise ratio.  If the local continuum level (computed above) is greater than 1.5 raw counts per 0.013 \AA\ pixel, we bin the spectrum by 8 pixels or 0.104 \AA.  A diffuse emission feature is about 3.5 binned pixels wide.  If the continuum level is less than 1.5, we bin by 14 pixels or 0.182 \AA, a value chosen so that a diffuse emission feature is spanned by 2 binned pixels.  Most of our emission features are broad enough that 14-pixel binning provides sufficient spectral resolution; however, we find that emission features whose Gaussian components have FWHM values less than 100 \kms\ can be undersampled at this resolution, so we lower the threshold for 8-pixel binning to 0.5 raw counts per 0.013 \AA\ pixel for spectra containing these narrow features.  The binning applied to each of our spectra is listed in Table \ref{tab_detections}.  After binning, the FLUX and ERROR arrays are converted from units of \flux\ to LU pixel$^{-1}$, and the ERROR array is smoothed by 5 pixels, which removes zero-valued error bars without significantly changing its shape.

To derive the line parameters of each emission feature, we fit model spectra to the flux-calibrated data using the nonlinear curve-fitting program \specfit\ \citep{Kriss:94}, which runs in the IRAF\footnote[2]{The Image Reduction and Analysis Facility (IRAF) is distributed by the National Optical Astronomy Observatories, which is supported by the Association of Universities for Research in Astronomy (AURA), Inc., under cooperative agreement with the National Science Foundation.} environment, to perform a $\chi^2$ minimization of the model parameters.  Our synthetic line profiles are described above.  The program interpolates between tabulated curves to reproduce the shape of the emission line.  Free parameters in the fit are the level and slope of the continuum (assumed linear) and the intensity, wavelength, and Gaussian FWHM of the model line.  For most sight lines, we model only the region 1028.7--1036.5 \AA.  We do not attempt to fit the \osix\ 1038 \AA\ feature, as it is only half as strong as the 1032 \AA\ line in an optically-thin gas and is often blended with emission from interstellar \ctwo * $\lambda 1037.02$.  

Best-fit values for the \osix\ $\lambda 1032$ line parameters are reported in Table \ref{tab_detections}, and plots of the spectra and best-fit models are presented in Figures \ref{fig_3sigma} and \ref{fig_2sigma}.  In the table and figures, as throughout this paper, we distinguish between  3\sig\ detections, for which $I/\sigma(I) \geq 3$, and 2\sig\ detections, for which $I/\sigma(I) < 3$.  All of these features meet our requirement of having an intensity greater than three times the uncertainty in the local continuum and are thus statistically significant; however, to derive the physical properties of the \osix -bearing gas, we consider only those features whose intensities are certain at the 3\sig\ level.  

The FWHM values in Table \ref{tab_detections} include the smoothing imparted by the instrument optics; values less than $\sim$ 25 \kms\ indicate that the emission does not fill the LWRS aperture.  In such cases, the surface brightness of the emitting region will be underestimated, because the conversion to LU assumes that the emission fills the aperture, and the velocity of the emitting gas will be uncertain, because it may not be centered in the aperture.

The quoted uncertainties are the error bars returned by \specfit, which are obtained from the error matrix and correspond to a 1\sig\ confidence interval for a single interesting parameter.  For multi-parameter fits with more than one interesting parameter, this method can underestimate the true uncertainty in each, so we have computed more rigorous error bars for sight line I20509 (for which the S/N is high and the \osix\ $\lambda 1032$ line is broad) in the following way:  For each model parameter, we begin with the best-fit value, then increase it, while re-optimizing the other model parameters, until $\chi^2$ increases by 1.0 \citep{Press:88}.  We find that the two methods yield identical error bars, suggesting that the \specfit\ errors are sufficient for well-sampled spectra.  

Both the \fuse\/ flux calibration, which is based on theoretical models of white-dwarf stellar atmospheres, and the solid angle of the LWRS aperture are known to within about 10\% \citep{Sahnow:00}.  Added in quadrature, they contribute a systematic uncertainty of $\sim$ 14\% \citep{Shelton:01} in addition to the statistical uncertainties quoted in Table \ref{tab_detections}. 

Our best-fit line intensities are lower limits, in the sense that the intrinsic \osix\ intensity may be higher than is observed.  We discuss the effects of dust extinction in \S \ref{sec_extinction}.  Resonance scattering within the \osix -bearing gas can also have a strong effect on the observed intensity.  For an optically thin gas, the intensity ratio $I_{1032} / I_{1038} = 2$, whereas an optically thick plasma yields a ratio of unity.  Since the \osix\ $\lambda 1038$ line is generally difficult to measure (as mentioned above) and its intensity has a large uncertainty, the resulting line ratio is usually inconclusive.  We therefore do not attempt to estimate the self absorption along our detection sight lines.

Because molecular hydrogen is ubiquitous in the ISM, we must consider the effects of \htwo\ absorption and emission on our results.
The \htwo\ features nearest \osix\ $\lambda 1032$ are the Lyman (6,0) $P(3)$ and $R(4)$ lines at 1031.19 and 1032.36 \AA, respectively, but these features are weak in cold ($\sim$ 100 K), diffuse clouds \citep{Shull:00} and will not significantly reduce the observed intensity of the \osix\ $\lambda 1032$ feature.
Another possibility is that fluorescent \htwo\ contributes to the observed  \osix\ emission.  Following \citet{Shelton:01}, we search for the Werner (0,1) $P(3)$ $\lambda 1058.82$ line, which for cool clouds bathed in ultraviolet light should be at least 50\% brighter than any of the \htwo\ emission lines between 1030 and 1040 \AA\ \citep{Sternberg:89}, but none of our 3\sig\ sight lines exhibits a statistically significant emission feature near 1059 \AA.

\subsection{Upper Limits on \osix\ Emission}

\begin{figure}
\figurenum{4}
\plotone{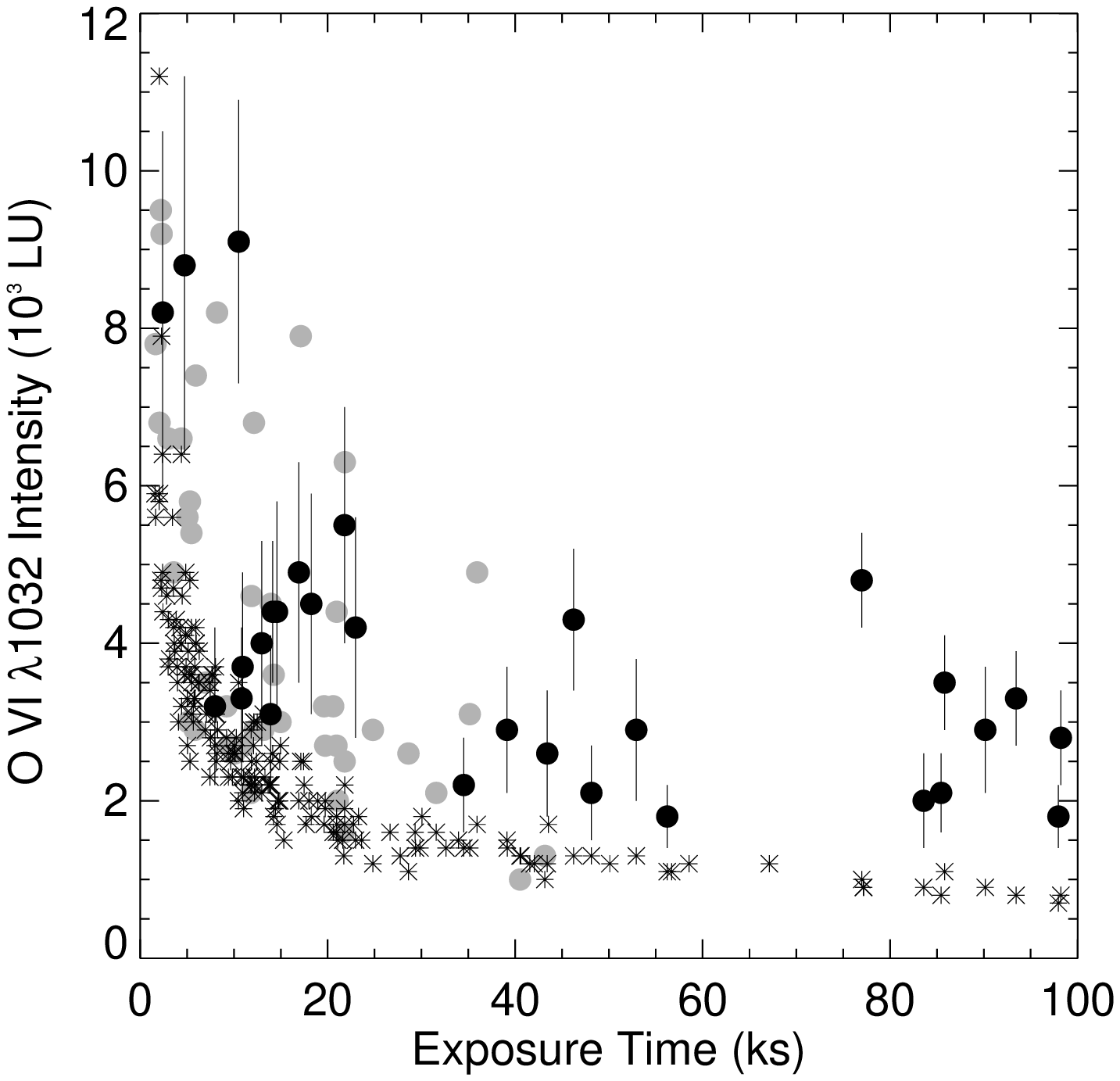}
\caption{Observed \osix\ $\lambda 1032$ intensities and upper limits.  Black circles represent 3\sig\ detections. Grey circles (without error bars) represent 2\sig\ detections.   Asterisks represent the 3\sig\ upper limits computed for all sight lines as described in the text.  \fuse\/ can detect \osix\ emission as faint as 2000 LU in 18 ks and as faint as 1000 LU in 80 ks (night only).}
\label{fig_limits}
\end{figure}

For non-detection sight lines, we compute 3\sig\ upper limits to the intensity of an \osix\ emission feature using the mean value of the FLUX array between 1030 and 1035 \AA\ and assuming a line width of 28 pixels (106 \kms).   The resulting limits are given in Table \ref{tab_limits}.  To obtain more data points with long integration times, we repeat the exercise using our detection sight lines, but masking out the \osix\ line when computing the continuum level.  Figure \ref{fig_limits} presents both sets of limits as a function of exposure time.  \fuse\/ can detect diffuse \osix\ emission as faint as 2000 LU in 18 ks of night exposure time and as faint as 1000 LU in $\sim$ 80 ks of night time.

\subsection{Comparison with Previously-Published Sight Lines\label{sec_published}}

It is instructive to compare our results with the previously-published values listed in Table \ref{tab_published}.  Four of the published sight lines, A11701, B00303, B00302, and P11003, fail one or more of our statistical filters.  A11701 yields only an upper limit on the \osix\ intensity, while the other three produce 2\sig\ detections.  In three of these four cases, the original analysis used both day and night data, which explains our lower S/N ratios.  Similarly, CalFUSE rejects most of the night exposure time for sight line S40561 due to limb-angle violations, raising our upper limit on the \osix\ intensity.  For sight line S40548, we obtain the published line intensity if we fix the Gaussian line width to be 75 \kms, but find the best-fit intensity and Gaussian FWHM to be a factor of two larger.  Sight line  B12901 is discussed in \S\ \ref{sec_B12901}.  The remaining sight lines,  A11703 and I20509, yield 3\sig\ detections in our survey.  In both cases, our intensities and derived FHWM values agree (within the errors) with the previously-published values.  In neither case do our LSR velocities agree, illustrating the difficulty of deriving an accurate wavelength scale from nearby airglow features, as previous authors were required to do.  

The results of \citet{Otte:06} are not included in Table \ref{tab_published}.  Comparison with their results is complicated by the fact that several of their sight lines appear in our survey with additional exposure time or combined with nearby lines of sight.  Even so, when they report a 3\sig\ detection, their best-fit line parameters generally agree with ours within the quoted errors.  The recent \spear\/ results are discussed in \S \ref{sec_emission}.

\section{Results\label{sec_results}}

Measured \osix\ $\lambda 1032$ intensities and upper limits for our survey sight lines are presented in \fig{fig_limits}.  
\osix\ emission is detected at 3\sig\ significance along 29 lines of sight.  
Measured intensities range from 1800 to 9100 LU, with a median of 3300 LU, 
an average of 3900 LU, and a standard deviation 1900 LU\@.  
An additional 44 sight lines exhibit \osix\ $\lambda 1032$ emission at lower significance, 
while 110 non-detection sight lines provide 3\sig\ upper limits on the \osix\ intensity,
35 of them less than 2000 LU\@.  The median value of all the 3\sig\ limits is 2600 LU\@.  
The upper limits are strongly correlated with exposure time and generally lower
than the measured intensities along detection sight lines with comparable exposure time.  
For our complete sample,
the \osix\ detection rate (at 3\sig\ significance) is 16\%.
If we consider only sight lines with exposure times greater than 18 ks
(for which 3\sig\ upper limits are less than 2000 LU), 
the detection rate rises to 30\%.  

\subsection{Dust Extinction\label{sec_extinction}}

\begin{figure}
\figurenum{5}
\plotone{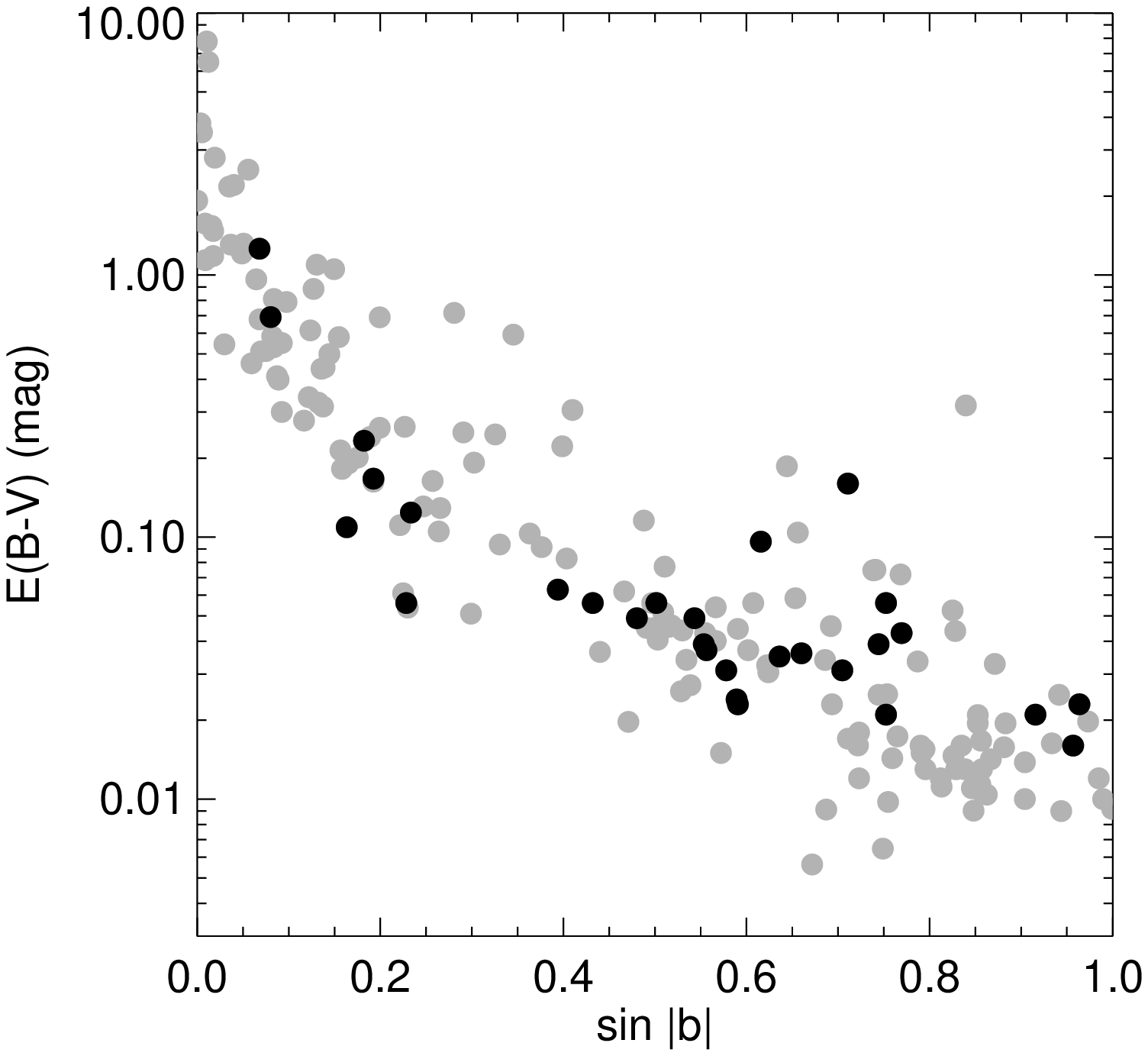}
\caption{Variation of color excess with Galactic latitude.  Black circles represent sight lines with 3\sig\ detections.  Grey circles represent the rest of our sample.  Data are from \citealt{Schlegel:98}.}
\label{fig_ebv_sinb}
\end{figure}

UV emission is strongly attenuated by interstellar extinction.
A color excess $E(B-V)$ of 0.05 magnitudes reduces the flux at 1032
\AA\ by a factor of 2, and a color excess of 0.18 magnitudes by a
factor of 10 \citep{Fitzpatrick:99}.  Thus, when we detect \osix\ emission in a direction
with high extinction, we assume that the emission arises in the
local ISM, \ie, closer than the dust causing most of the extinction.
On the other hand, emission detected in directions of low extinction
does not necessarily arise in the distant ISM, but may come from
relatively nearby gas.  Figure \ref{fig_ebv_sinb} presents a plot of color excess versus
$\sin |b|$, where $b$ is the Galactic latitude, for each sight line in our survey.  
We see that extinction
falls rapidly as one moves away from the Galactic plane.  We caution that
interstellar extinction is variable on small spatial scales,
and the \citet*{Schlegel:98} values presented in \fig{fig_ebv_sinb}
are based on \iras\/ maps with low spatial resolution.  An example
of an \osix\ detection through a region of patchy extinction is
discussed in \S\ \ref{sec_B12901}.   

\subsection{\osix\ Emission-Line Intensities\label{sec_emission}}

\begin{figure}
\figurenum{6}
\plotone{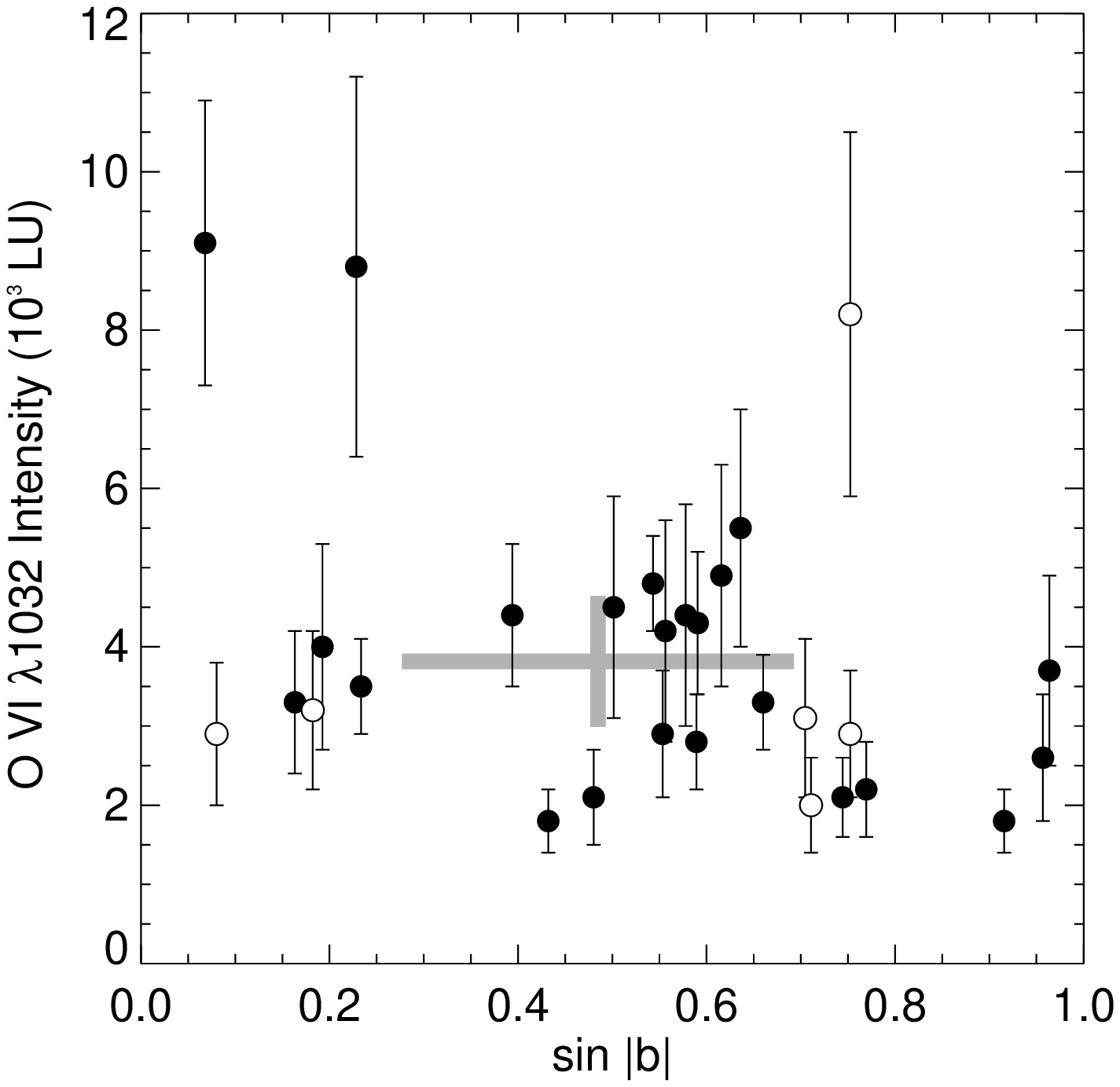}
\caption{Variation of observed \osix\ $\lambda 1032$ intensity with Galactic latitude.  Only 3\sig\ detections are shown; open circles represent those with $|v_{\rm LSR}| > 120$ \kms.  The grey cross represents the \spear\ observation of the 15\degr -radius region centered on the north ecliptic pole ($l = 123,\; b = +29\degr$) reported by \citealt{Korpela:06}.}
\label{fig_intensity_sinb}
\end{figure}

In \fig{fig_intensity_sinb}, we plot \osix\ intensity against $\sin |b|$, where
$b$ is the Galactic latitude of the observed sight line.  
The emission features plotted as open circles have
absolute velocities greater than 120 \kms.  We discuss these sight lines
and their possible relationship with high-velocity clouds (HVCs)
in \S\ \ref{sec_velocity}.  
Of the 23 low-velocity sight lines, two have \osix\ intensities
greater than 8000 LU, while the others range from 1800 to 5500 LU.  
Both high-intensity sight lines probe regions known to be energized by hot stars and supernova remnants. 
P12011 (discussed in \S\ \ref{sec_Jupiter}) intersects the Monogem Ring, which is about 800 pc away, and S50508 probes the outskirts of the Vela supernova remnant (250 pc distant) and the Gum Nebula, which probably lies somewhat beyond Vela.  

Based on observations with \spear, \citet{Korpela:06} report a combined \osix\ $\lambda \lambda 1032, 1038$ intensity of $5724 \pm 570$ (statistical) $\pm\, 1100$ (systematic) LU for the region of 15\degr\ radius centered on the north ecliptic pole ($l = 123,\; b = +29\degr$).  The grey cross in \fig{fig_intensity_sinb} represents the \spear\ sight line.  Its intensity is consistent with the \fuse\/ data points in this latitude range.  Two of our 3\sig\ detections and one of our strong upper limits lie within the region sampled by \spear: P10429 ($I(1032) = 2900 \pm 800$ LU),  Z90715 ($I(1032) = 4400 \pm 1400$ LU), and D11701 ($I(1032) < 1400$ LU).

Restricting our attention to the low-velocity, low-intensity \osix\
measurements plotted in \fig{fig_intensity_sinb}, 
we note that the sight lines at high latitude tend to be
fainter than average.  The median intensity for this sample of 21
detections is 3500 LU, with a 95\% confidence interval of (2700, 4500) LU.
The median intensity for sight lines with $\sin |b| < 0.7$ is 4100 LU with
confidence interval (3100, 4400) LU.  For sight lines with $\sin |b| >
0.7$, these values are 2200 LU, (1800, 3700) LU.  The median estimates
for the two groups are quite different, but the 95\% confidence
intervals overlap.  The overlap is driven by a single high-latitude
point (A11703, at 3700 LU), which determines the upper limit of the
confidence interval for its group.  Although the equality of medians
cannot be statistically excluded (at the 95\% level), the data suggest
that \osix\ emission tends to be fainter at high than at low latitudes.

A variation in the \osix\ intensity with latitude might result from either of two competing effects:  
First, interstellar extinction falls steeply as one moves off the Galactic plane (\fig{fig_ebv_sinb}), so attenuation is lower at high latitudes.  
Second, the path length through the Galactic disk scales inversely with $\sin |b|$, so high-latitude sight lines may intersect fewer \osix -bearing regions.  
We have calculated the relative importance of these effects for a uniform distribution of \osix -bearing clouds in the disk, assuming that the extinction is local.  
At low latitudes ($\sin |b| < 0.1$), attenuation dominates completely, and the observed emission must come from nearby regions.  
Intensities rise gradually for $0.1 < \sin |b| < 0.4$, as the extinction decreases faster than the path length.
At high latitudes, the two effects nearly cancel, and intensities are roughly constant for $\sin |b| > 0.4$.
Models assuming an exponential distribution of the \osix -bearing clouds yield similar results.
Thus, the low \osix\ intensities seen along sight lines with $\sin |b| > 0.7$ in \fig{fig_intensity_sinb} cannot be explained only by differences in path length.
We suggest that the \osix -emitting regions at high latitudes are intrinsically fainter than those at low latitudes.  
The faint regions likely constitute a population in the thick disk or halo (perhaps with properties similar to those of the HVCs located in the same region of \fig{fig_intensity_sinb}), while the brighter regions lie in the disk of the Galaxy.  It is possible that the \osix\ detected along mid-latitude
sight lines includes both disk and thick-disk emission, but for the lowest-latitude sight lines, the emitting regions necessarily lie in the disk.

\subsection{\osix\ Emission-Line Velocities\label{sec_velocity}}

The six 3\sig\ detections plotted as open circles in \fig{fig_intensity_sinb} have absolute velocities greater than 120 \kms, corresponding to the velocities of HVCs.  \osix\ absorption associated with \hone\ high-velocity clouds was first detected by \citet{Sembach:00} and \citet{Murphy:00}.  \citet{Savage:03} report that, when a known \hone\ HVC is present along a line of sight to an object, high-velocity \osix\ absorption spanning the approximate velocity range of the \hone\ HVC is usually seen.  The association of \osix\ with HVCs suggests that the \osix\ may be produced at interfaces between the \hone\ clouds and hot, low-density gas in the Galactic corona or Local Group.  

Two of our high-velocity sight lines, B12901 and S40549, intersect known \hone\ HVCs and share the clouds' velocities.  B12901 intersects the Magellanic Stream, while S40549 probes the HVC known as Complex C (see Fig.\ 16 of \citealt{Wakker:03} and Figs.\ 11 and 13 of \citealt{Sembach:03}).  These sight lines are discussed in \S \ref{sec_hvc}.
Sight line S40543, which is near S40549 on the sky, also intersects Complex C.  The emission in S40543 has a high positive velocity, while Complex C has a high negative velocity; however, high-positive-velocity \osix\ absorption is seen nearby \citep{Sembach:03}.
Sight lines C06401 and S30402 do not intersect HVCs.  
P10414 and S30402 probe low-latitude sight lines with high extinction, so their emission probably originates in the nearby disk (but see the discussion of patchy extinction in \S\ \ref{sec_B12901}).  Perhaps they sample fast-moving gas in previously undetected supernova remnants.  (Note that the \halpha\ intensity along S30402 is unusually high.)

\begin{figure}
\figurenum{7}
\plotone{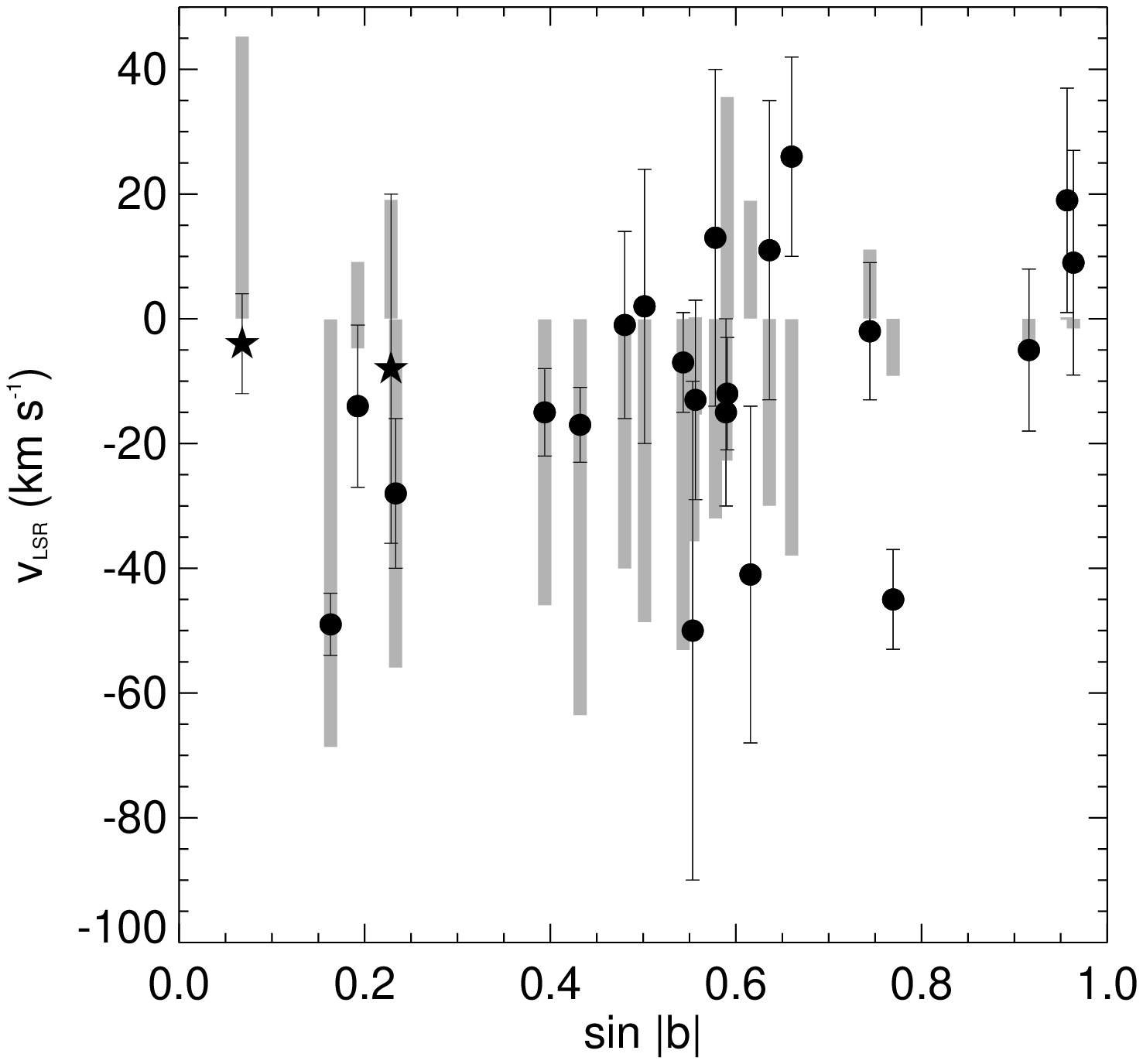}
\caption{Variation of $v_{\rm LSR}$ with Galactic latitude.  Only 3\sig\ detections are shown.  The two high-intensity sight lines discussed in \S \ref{sec_emission} are plotted as stars.  Sight lines with $|v_{\rm LSR}| > 120$ \kms\ are not included.  Grey bars represent the range of LSR velocities predicted for the first 5 kpc along each sight line by a simple model of Galactic rotation.}
\label{fig_velocity}
\end{figure}

In \fig{fig_velocity}, the LSR velocities of our low-velocity 3\sig\ emission features 
are plotted against $\sin |b|$.  
The grey bars represent the range of LSR velocities predicted 
for each sight line by a simple model of Galactic rotation.
Assuming a differentially-rotating halo with a constant velocity
of $v = 220$ \kms, we compute the expected radial velocity as a
function of distance for the first 5 kpc along each line of sight.  
The model ignores broadening due to turbulence or any other motion.
The three sight lines with $\sin |b| > 0.5$ and 
$v_{\rm LSR} < -40$ \kms\ may probe intermediate-velocity clouds.   
The velocities of the remaining sight lines are generally consistent with Galactic rotation. 

\subsection{\osix\ Emission-Line Widths\label{sec_FWHM}}

Seven of our 3\sig\ sight lines have best-fit Gaussian FWHM values less than 25 \kms, which indicates that their \osix -emitting regions do not fill the $30\arcsec \times 30\arcsec$ LWRS aperture.  Of the seven, C07601 and S50504 include data from lines of sight separated by an arc minute or more.  The other five each sample a single line of sight.  The data set with the longest exposure time, and thus the highest S/N, is S50510.  We fit its spectrum with a model similar to that used to parameterize the \oone\ $\lambda 1039$ airglow line: a top-hat function convolved with a Gaussian.  The width of the top hat, together with the line intensity and centroid, are free parameters in the fit, but the Gaussian FWHM is fixed at 15 \kms, the instrumental resolution for a point source.  (If allowed to vary freely, the Gaussian FWHM falls below 1 \kms.)  For numerical simplicity, we fit the raw-counts spectrum, first binning the data by four pixels, which the raises the background to $\sim$ 10 counts per binned pixel.  The width of the best-fit top-hat function is $80 \pm 15$ \kms, significantly less than the 106 \kms\ width of an aperture-filling emission feature.  By simple scaling, we derive an angular size of about 23\arcsec\ for the emitting region.  Its spatial scale is distance dependent; at a distance of 1 kpc, 23\arcsec\ corresponds to about 0.1 pc.

\subsection{Comparison with H$\alpha$ and Soft X-ray Emission\label{sec_sxr}}

Perusal of the H$\alpha$ map produced by the Wisconsin H-Alpha Mapper (WHAM) Northern Sky Survey \citep{Haffner:03} reveals that our \osix\ detection sight lines probe a variety of environments: toward \ion{H}{2} regions, filaments, and bubbles, as well as toward faint, featureless, ionized gas.  
For each of our sight lines, Tables \ref{tab_detections} and \ref{tab_limits} list the H$\alpha$ intensities (integrated over one degree on the sky and over the velocity range $-80$ \kms\ $< v <$ +80 \kms) measured by WHAM.  
We find no correlation between the \osix\ and H$\alpha$ intensities.

\begin{figure}
\figurenum{8}
\plotone{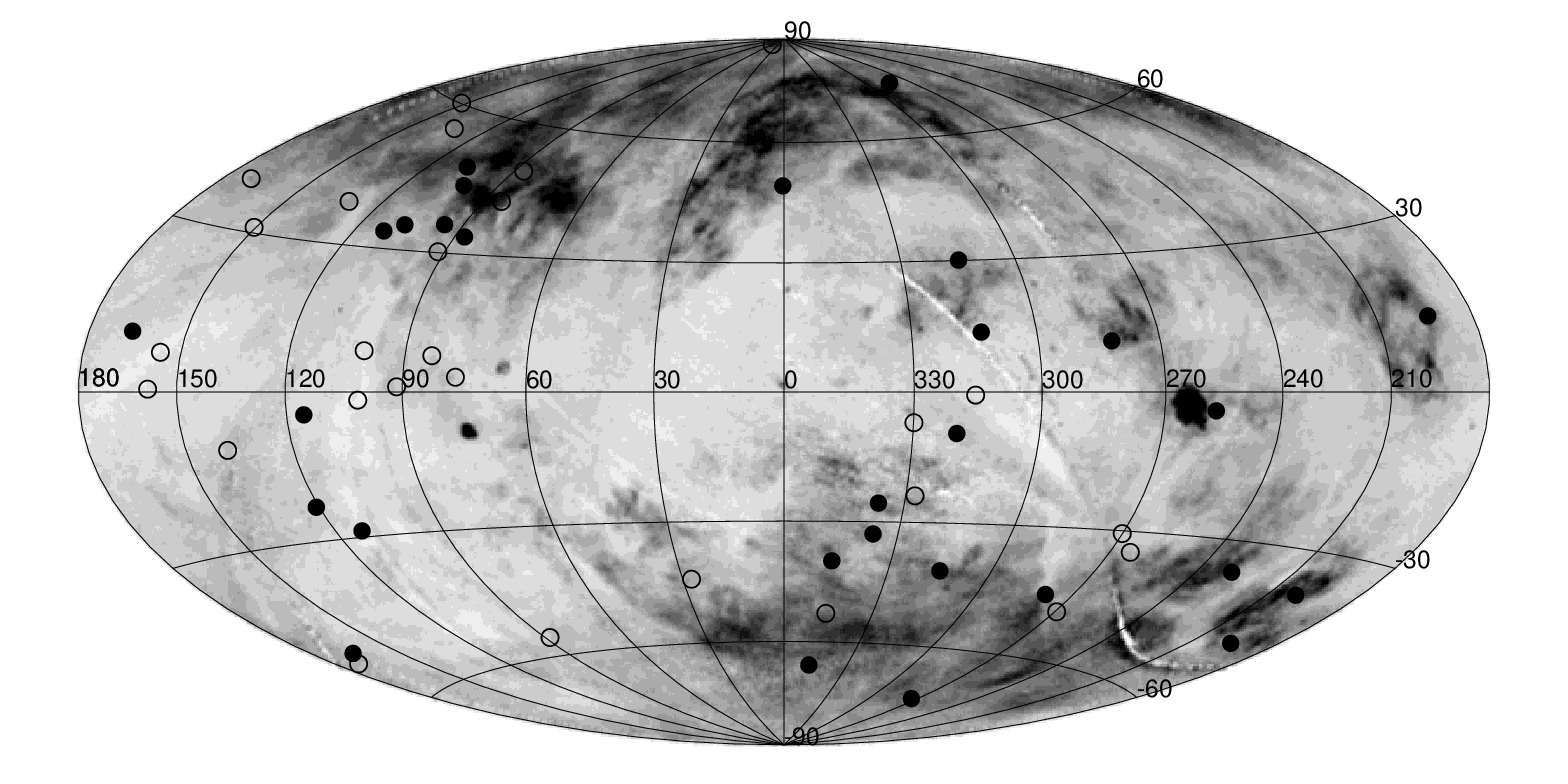}
\caption{Distribution of survey sight lines on the SXR sky. Solid circles represent 3\sig\ detections.  Open circles represent non-detections with upper limits less than 2000 LU.  The gray-scale map shows the 1/4 keV X-ray sky observed by \rosat\/ \citep{Snowden:97}.  Galactic coordinates are used in a Hammer-Aitoff projection.}
\label{fig_sxrmap}
\end{figure}

One possible origin of the observed \osix\ emission is hot gas cooling from temperatures of $10^6$ K or more.  Gas at these temperatures is observable in the soft X-ray (SXR) regime.  Figure \ref{fig_sxrmap} presents the distribution of our 3\sig\ detections and strong upper limits on a map of the 1/4 keV X-ray sky observed with \rosat\/ \citep{Snowden:97}.  Prominent features include the North Polar Spur, which arches from $(l,b) = (30, +30)$ to $(290, +60)$ and may be probed by sight line A11703 (284, +75); the Vela supernova remnant at $(260, -5$), whose outer regions may be probed by sight line S50508 $(257, -4)$; and the Monogem Ring, a supernova remnant centered near $(205, +15)$, which may be probed by sight line P12011 $(194, +13)$.
Tables \ref{tab_detections} and \ref{tab_limits} list the \rosat\/ 1/4 keV SXR emission observed toward each of our sight lines.  
We find no correlation between the \osix\ and SXR intensities.  In their \osix\ absorption-line survey, \citet{Savage:03} find no significant correlation between $N($\osix) and either $I({\rm H}_{\alpha})$ or $I($SXR).

\section{Properties of the \osix -Bearing Gas\label{sec_discussion}}

Measurements of \osix\ emission and absorption can be combined to provide valuable diagnostics of the \osix -bearing gas, so long as both probe the same interface or transition region.
For a particular region, the intensity scales as $n_e^2 L$, where $n_e$ is the electron density and $L$ the path length through the region, while the column density scales as $n_e L$.  
From the ratio $I($\osix$)/N($\osix), we can derive the electron density of the plasma \citep{Shull:Slavin:94} and, assuming an oxygen abundance, the path length through it.  
This calculation assumes that the density of the region is uniform.  Despite this simplification, our results should be correct to within an order of magnitude.

A typical sight line through the Galaxy could intersect several \osix\ regions, each of which would contribute differently to the total \osix\ column density and intensity.  
Because we lack the spectral resolution to isolate the contributions of individual regions, we must identify cases in which the integrated \osix\ column density and intensity can be attributed to a single region.  The likelihood that absorption and emission sight lines probe the same \osix\ region is highest for nearby stars: at a distance of 100 pc, the  3\farcm5 offset between the \fuse\/ LWRS and MDRS apertures corresponds to $\sim$ 0.1 pc.  Moreover, sight lines to nearby stars are more likely to intersect only a single interface or transition region than sight lines to distant objects.  We discuss one such sight line in \S \ref{sec_white_dwarfs}.  For more distant emitting regions, such as the HVCs discussed in \S \ref{sec_hvc}, we equate the absorbing and emitting gas based on their velocities and consider the range of reported \osix\ column densities for the HVC.  

\subsection{\osix -Bearing Gas in the Galactic Disk\label{sec_white_dwarfs}}

Four of our 3\sig\ sight lines correspond to nearby white dwarfs in the \osix\ absorption-line survey of \citet{Savage:06}.  Their \osix\ absorption-line measurements and our \osix\ emission-line velocities are presented in Table \ref{tab_lehner}.  (Note that the velocities in Table \ref{tab_lehner} are heliocentric.)
The best case in our sample for combining emission and absorption measurements is sight line P20411.  The velocity of its \ion{O}{6} emission ($v_{\rm Helio} = -6 \pm 15$ \kms) is consistent with that of the \osix\ absorption \citep[$v_{\rm Helio} = -3.8\pm3.6$ \kms;][]{Savage:06} measured toward WD\,0004$+$330 (GD~2), a DA white dwarf at a distance of 97 pc \citep{Vennes:97}.
The relative narrowness of the \osix\ emission feature (the best-fit value of the intrinsic Gaussian FWHM is $60 \pm 40$ \kms) suggests that it probes a single emitting region.
P20411 was observed with the MDRS aperture centered on the star, so the LWRS aperture probes a sight line passing within $\sim$ 0.1 pc of the white dwarf.  

The spectrum of GD~2 shows absorption from molecular hydrogen with a column density $\log N($\htwo) = 14.46 \citep[cm$^{-2}$;][]{Lehner:03}.  Since \htwo\ is assumed to form on dust grains, we must consider the possibility of dust extinction along this line of sight.  To estimate the extinction toward this star, we compare it with HZ~43, another nearby DA white dwarf that shows no \htwo\ absorption.  \citet*{Finley:97} derive effective temperatures of 49,360 and 50,822 K for GD~2 and HZ~43, respectively.  If their temperatures are nearly equal, then any difference in their \bv\ colors is likely due to reddening toward GD~2.  These colors are $-0.29$ and $-0.31$ magnitudes for GD~2 and HZ~43, respectively \citep{Eggen:68, Bohlin:01}, meaning that \ebv\ = 0.02 toward  GD~2, which corresponds to an attenuation of 30\% at 1032 \AA\ \citep{Fitzpatrick:99}.  More recent analyses suggest that GD~2 is somewhat cooler than HZ~43 (for example, \citealt{Barstow:03} derive temperatures of 45,460 and 46,196 K from the star's Balmer and Lyman lines, respectively), which would explain the color difference without invoking dust, so we will treat this reddening as an upper limit.

The observed \ion{O}{6} $\lambda$1032 intensity is 2100 LU and the absorbing column is $N$(\osix) = $6.2 \times 10^{12}$ cm$^{-2}$. 
We assume that the intrinsic intensity of the \ion{O}{6} $\lambda$1038 emission is half that of the 1032\,\AA\ line, as would be expected if the gas were optically thin.  
Assuming a temperature of $2.8\times10^5$\,K, Equation 5 from \citet{Shull:Slavin:94} yields an electron density $n_{\rm e}=0.22$ cm$^{-3}$ if \ebv\ = 0.00 and $n_{\rm e}=0.29$ cm$^{-3}$ if \ebv\ = 0.02. 
The \ion{O}{6} absorption line has a Doppler parameter $b \sim 30$ \kms; if thermal, it implies a temperature of $4.2 \times 10^5$ K, which does not change the derived electron density.

To calculate the \ion{O}{6} density and the path length through the emitting
region, we need the oxygen abundance and the fraction of oxygen in O$^{+5}$. 
\citet{Oliveira:05} derive a mean O/H ratio for the Local Bubble of $(3.45 \pm 0.19) \times 10^{-4}$. 
For plasmas in collisional ionization equilibrium, the O$^{+5}$ fraction peaks at
22\% when the gas temperature is $2.8 \times 10^5$ K \citep{Sutherland:Dopita93}. 
With these values, and
assuming that the gas is completely ionized ($n_{\rm e} = 1.2 n_{\rm H}$), we
derive an \ion{O}{6} density of $1.4 \times 10^{-5}$ cm$^{-3}$ and a path length
through the gas of $\sim$ $4.4 \times 10^{17}$ cm or 0.14 pc for zero reddening.  
If \ebv\ = 0.02, the \ion{O}{6} density and the path length through the gas become $1.8 \times 10^{-5}$ cm$^{-3}$ and 0.11 pc, respectively.

\citet{Savage:06} report that the \osix\ absorption feature in the spectrum of GD~2 is well detected (4.8\sig) and closely aligned in velocity with the \ion{C}{2} and \ion{O}{1} absorption, consistent with \osix\ formation in a condensing interface between the cool gas traced by the \ion{C}{2} and \ion{O}{1} and a hot exterior gas.
\citet{Bohringer:87} model conductive interfaces around spherical interstellar clouds embedded in a hot interstellar medium.  One of their models (model H) assumes a cloud of radius $3 \times 10^{17}$ cm in an external medium of temperature $5 \times 10^5$ K.  The model predicts a particle density at the cloud surface of $n_0 = 0.73$ cm$^{-3}$ (our mean cloud density is $n = 0.40$ cm$^{-3}$), a column density through the interface region of $N$(\osix) = $3.2 \times 10^{12}$ cm$^{-2}$, and a mean temperature for the O$^{+5}$ ions of $4.7 \times 10^5$ K.  These predictions are within a factor of 2 of our derived cloud parameters.  The high temperature is a general feature of the \citeauthor{Bohringer:87} models.

Another white-dwarf sight line worthy of comment is P10411 (WD\,0455$-$282).  While the velocities of its principal \osix\ emission and absorption components disagree, \citet*{Holberg:98} report \ion{Si}{4} and \ion{C}{4} absorption in \iue\/ spectra of this star at $v_{\rm Helio} = 16.21 \pm 2.66$ \kms, and \citet{Savage:06} report weak \osix\ absorption at the same velocity.  \citeauthor{Holberg:98} argue that this absorption is circumstellar, rather than interstellar.  If so, the emission that we observe at $v_{\rm Helio} = 7 \pm 9$ \kms\ may come from this circumstellar material.  With an effective temperature $T_{\rm eff}$ = 57,200 K, the white dwarf is too cool to produce O$^{+5}$ through photoionization, so the emission must be powered by shocks in the circumstellar material, perhaps generated by the interaction of material from previous episodes of mass loss.  This mechanism is at work in planetary nebulae \citep{Villaver:02}, and high-ionization lines have been observed in the spectra of low-$T_{\rm eff}$ central stars of planetary nebulae (J.~Herald, private communication).

\subsection{\osix -Bearing Gas in High-Velocity Clouds\label{sec_hvc}}

Two of our high-velocity sight lines, B12901 and S40549, intersect known \hone\ HVCs and share the clouds' velocities (\S\ \ref{sec_velocity}).  Data set B12901 consists of spectra obtained along three closely-spaced sight lines that probe the Magellanic Stream, but only one of them, I2050501/I2050510, exhibits significant \osix\ $\lambda 1032$ emission (\S\ \ref{sec_B12901}).  Its LSR velocity is $206 \pm 13$ \kms, and its intensity is $3000 \pm 600$ LU  (Table \ref{tab_b12901}).  The \osix\ column densities of HVCs associated with the positive-velocity portion of the Magellanic Stream range from $\log N$ = 13.78 to 14.33.  Their velocities range from 183 to 330 \kms\ with a mean of 232 \kms\ \citep{Sembach:03}.  Combining these absorption and emission measurements, we derive an electron density for the \osix -bearing gas of 0.01--0.03 cm$^{-3}$.  Adopting the LMC oxygen abundance \citep[$2.24 \times 10^{-4}$ O atoms per H atom;][]{Russell:92}, we find a path length through the emitting gas of 16--200 pc.

Sight line S40549, with $v_{\rm LSR} = -172 \pm 9$ \kms\ and $I(1032) = 8200 \pm 2300$ LU, probes Complex C, a large assembly of high-velocity ($-170 \la v_{\rm LSR}  \la -100$ \kms) \hone\ in the northern Galactic sky between $l \sim 30\degr$ and $l \sim 150\degr$.  Measured \osix\ column densities for sight lines through Complex C range from $\log N$ = 13.67 to 14.22 \citep{Sembach:03}.  From these values, we derive an electron density for the \osix -bearing gas of 0.03--0.11 cm$^{-3}$.  Assuming a Galactic O/H ratio of $6.61 \times 10^{-4}$ \citep{Allen:73}, we find a path length through the emitting gas of 1.1--14 pc.  The color excess toward S40549 is \ebv\ = 0.02 (Table \ref{tab_detections}), which attenuates emission at 1032 \AA\ by 30\% \citep{Fitzpatrick:99}.  Correcting for this attenuation raises the electron density and reduces the path length through the emitting gas by the same factor.  We expect the extinction along sight line B12901 to be similar, as the reddening in this region is patchy and quite low along nearby sight lines (\S\ \ref{sec_B12901}).

The emitting regions probed by sight lines B12901 and S40549 appear to span an order of magnitude in $n_e$ and two orders of magnitude in path length.  While this spread may reflect real variations in the properties of thick disk/halo gas, 
we should note that S40549 is one of the shortest exposures in our sample.  Its \osix\ $\lambda 1032$ feature is both extremely narrow (FWHM = $2 \pm 20$ \kms) and unusually bright.  Careful analysis confirms that this feature is statistically significant at the 3\sig\ level; however, additional exposure time would be helpful to confirm this result.

\subsection{Limits on the Gas Density\label{sec_heckman}}

Along sight lines through the Galactic disk, 
most of the \osix\ column densities measured to date lie in the range
$\log N($\osix$) = 12.6$ to 14.0, 
and along sight lines through the Galactic halo, in the range
$\log N($\osix$) = 13.7$ to 14.7 \citep{Savage:03}.  
Excluding emission associated with known SNRs and the short S40549
exposure discussed in \S\ \ref{sec_hvc}, our \osix\ $\lambda 1032$ intensities 
also span a narrow range, from 1800 to 5500 LU.
(The lower bound reflects our sensitivity limits.)
The restricted range of observed column densities, together with
the narrow range of our measured intensities, suggests that the
volume densities of the \osix -bearing gas are likewise limited.
We have found two flavors of \osix -bearing gas: 
narrow interfaces in the Galactic disk with densities of about 0.1 cm$^{-3}$ and 
more extended cooling regions in the Galactic halo with densities of about 0.01 cm$^{-3}$.  
We argue that, in general, 
the volume densities of \osix -bearing gas in the Galactic disk and halo 
are unlikely to differ significantly from our derived values. 
Indeed, the observed range of \osix\ intensities is consistent with a constant volume density
in each environment, with \osix\ column density as the only variable.
In particular, our observations rule out the general presence of \osix -bearing 
thermal interfaces or cooling regions with densities of 1 to 10 cm$^{-3}$ or greater.

Our observations do not rule out the presence of hot, low-density gas with
significant \osix\ column densities.  \citet{Heckman:02} have shown that,
for radiatively cooling gas, the \osix\ column density at a temperature
of $2 \times 10^6$ K is comparable with that at $3 \times 10^5$ K because,
though the \osix\ fraction is smaller at the higher temperature,
the cooling times are much longer.  The \osix\ emission from such regions would
be too weak for detection by \fuse\/ because their emission measure would be too low.

\section{Summary\label{sec_summary}}

We have conducted a survey of diffuse \osix\ $\lambda 1032$ emission in the Galaxy using archival data from the {\it Far Ultraviolet Spectroscopic Explorer (FUSE).}  Of our 183 sight lines, 29 show \osix\ emission at 3\sig\ significance.  Measured intensities range from 1800 to 9100 LU, with a median of 3300 LU.  An additional 35 sight lines provide upper limits of 2000 LU or less.
Though the presence of \osix\ emission along low-latitude, high-extinction sight lines suggests that these emitting regions are nearby (probably within a few hundred parsecs), other emitting regions are more likely to be associated with HVCs in the Galactic halo.

Analysis of 21 low-velocity, low-intensity, 3\sig\ emission features reveals that
the \osix -emitting regions at high latitudes are intrinsically fainter than those at low latitudes and may represent a distinct population of emitters.  
Line velocities are generally consistent with a simple model of Galactic rotation.
Some of the \osix -emitting regions appear to have angular sizes smaller than the \fuse\/ LWRS aperture, which places a distance-dependent constraint on their physical size.
By combining emission and absorption measurements through the same \osix -bearing regions, we find evidence for relatively narrow, high-density conductive interfaces in the local ISM and more extended, low-density regions in HVCs.  
Based on the narrow range of \osix\ intensities in our sample and of \osix\ column densities in the Galactic disk and halo, we argue that the volume densities of \osix -bearing regions in each environment are unlikely to differ significantly from our derived values. 

\acknowledgments

The authors thank Ricardo Velez for his assistance with the initial data reduction for this project.
We acknowledge with gratitude the ongoing efforts of the \fuse\/ P.I.\@ team to make this mission successful.  R.~S. thanks J.~Cuervo for interesting discussions about statistical
methods and for help with their implementation in STATA.
This research has made use of the NASA/IPAC Extragalactic Database (NED), which is operated by the Jet Propulsion Laboratory, California Institute of Technology, under contract with the National Aeronautics and Space Administration; the NASA Astrophysics Data System (ADS); and the Multimission Archive at the Space Telescope Science Institute (MAST).  STScI is operated by the Association of Universities for Research in Astronomy, Inc., under NASA contract NAS5-26555. Support for MAST for non-HST data is provided by the NASA Office of Space Science via grant NAG5-7584 and by other grants and contracts.
This work is supported by NASA grant NAS5-32985 to the Johns Hopkins University.


\begin{appendix}

\section{Notes on Individual Lines of Sight}

\subsection{P12011 and P12012 (Jupiter)\label{sec_Jupiter}}

Our sample includes two observations of Jupiter (P12011 and P12012) designed to search for HD fluorescently pumped by solar Lyman $\beta$ emission.  In both cases, the MDRS aperture was centered on the planet, and the LWRS aperture was offset by 3.5 arcmin (or 10.3 Jupiter radii) in a direction perpendicular to the plane of the Jovian system.  No \osix\ emission is present in the P12012 spectrum ($\alpha$ =  07:06:57.55, $\delta$ = +22:24:10.1, J2000), but a 3\sig\ feature is seen in P12011 ($\alpha$ = 07:06:36.64, $\delta$ = +22:24:32.0).  The heliocentric velocity of the P12011 feature is $-0.2 \pm 25$ \kms, suggesting that the emission may have a local origin.  If it were scattered solar \osix\ emission, we would expect to see emission from \cthree\ $\lambda 977$ as well, but none is present in the SiC~2A spectrum from this observation.  Another possibility is that the observed emission is due to \htwo\ fluorescence, but comet spectra that show strong \htwo\ fluorescence near 1032 \AA\ also exhibit a strong 1163.7 \AA\ line (P.~D.\ Feldman 2005, private communication), which is not seen in the LiF~1B spectrum from this observation.  We conclude that the observed \osix\ is of interstellar origin.

\subsection{B12901\label{sec_B12901}}

The \fuse\/ observations of sight line B12901 were originally designed as a shadowing experiment to search for \osix\ emission arising in the Local Bubble \citep{Shelton:03}.  This sight line intersects a diffuse interstellar filament at a distance of $230 \pm 30$ pc \citep{Penprase:98}, well beyond the $\sim$ 100 pc radius of the Local Bubble \citep{Snowden:98}.   As its mean color excess is $E(B-V) = 0.17 \pm 0.05$ \citep{Penprase:98}, it was assumed that any \osix\ emission would have to come from material closer than the filament, presumably the Local Bubble itself.  

We detect an \osix\ $\lambda 1032$ emission feature of some 3\sig\ significance in this spectrum,  but with a velocity $v_{\rm LSR} = 192 \pm 19$ \kms, the emission is unlikely to originate in either the Local Bubble or the intervening filament.  \osix\ $\lambda 1032$ absorption features in the spectra of nearby white dwarfs exhibit velocities $|v| < 40$ \kms\ \citep[relative to the ISM as defined by the \ctwo\ $\lambda 1036$ line;][]{Oegerle:05}, and neutral gas in the filament moves at $0 < v_{\rm LSR} < 10$ \kms\ \citep{Penprase:98}.  \citeauthor{Penprase:98}\@ find that the filament is patchy, with $E(B-V) \leq 0.01$ toward some background stars.  Maps of the high-velocity \hone\ sky published by \citet{Sembach:03} and \citet{Wakker:03} reveal that sight line B12901 probes an HVC with $v_{\rm LSR} \sim 200$ \kms\ that is associated with the Magellanic Stream.  We conclude that the observed emission is produced by \osix -bearing gas associated with the HVC.

Data set B12901 consists of five observations (Table \ref{tab_data}) along three closely-spaced sight lines.  Since the filament is patchy, we searched for \osix\ emission along each sight line separately, using the technique described in \S\ \ref{SEC_MEASUREMENTS}.  Our results are presented in Table \ref{tab_b12901}: we find a strong \osix\ $\lambda 1032$ feature in the combined  I2050501/I2050510 spectrum and can set only upper limits on emission in the I2050601 and combined B1290101/B1290102 spectra.  The two I205 sight lines are separated by approximately 30\arcsec.  The strong variation in \osix\ intensity over such small angular scales is consistent with a patchy distribution of extinction in the filament.

In her original analysis, \citet{Shelton:03} found no evidence for \osix\ emission in these data and placed a 2\sig\ upper limit of 530 LU on the intensity of the \osix\ $\lambda 1032$ line.  Excluding the observed \osix\ $\lambda 1032$ emission feature, we derive a 2\sig\ upper limit of 600 LU for the full B12901 data set.  Shelton's conclusions about the physical conditions within the Local Bubble are therefore unchanged by our result.

\end{appendix}


\bibliographystyle{apj}
\bibliography{apjmnemonic,myref,h2ref,osix}



\clearpage
\LongTables
\tabletypesize{\tiny}
\begin{deluxetable}{ll}
\tablecolumns{2}
\tablewidth{0pt}
\tablecaption{\fuse\/ Observations Contributing to Each Survey Sight Line\label{tab_data}}
\tablehead{
\colhead{Sight Line} & \colhead{\fuse\/ Observation ID's}
}
\startdata
A01002 & {A0100201} \\
A03406 & {A0340601} \\
A03407 & {A0340701} \\
A04604 & {A0460404} \\
A04802 & {A0480202} \\
A05101 & {A0510102} \\
A09401 & {A0940102} \\
A10001 & {A1000101} \\
A11701 & {A1170101} \\
A11703 & {A1170303} {A1170404} \\
A13902 & {A1390201} {A1390202} {A1390203} {A1390204} {A1390205} \\
B00302 & {B0030201} \\
B00303 & {B0030301} \\
B01802 & {B0180201} \\
B01805 & {B0180501} {B0180502} \\
B04604 & {B0460401} {B0460501} \\
B06801 & {B0680101} \\
B12901 & {B1290101} {B1290102} {I2050501} {I2050510} {I2050601} \\
C02201 & {C0220101} {C0220102} \\
C02301 & {C0230101} \\
C03701 & {C0370101} \\
C03702 & {C0370201} {C0370202} {C0370203} {C0370204} {C0370205} {C0370206} \\
C06401 & {C0640101} {C0640201} {C0640301} \\
C07601 & {C0760101} {C0760201} {C0760301} {C0760401} \\
C10101 & {C1010101} {C1010201} {C1010301} {C1010302} {S4058501} \\
C10503 & {C1050301} {C1050302} {C1050303} {C1050304} \\
C11602 & {C1160202} \\
C11603 & {C1160301} {C1160402} \\
C16501 & {C1650101} {C1650102} {C1650103} {C1650104} {C1650105} {C1650106} {C1650107} {P1040701} \\
D04001 & {D0400101} \\
D05801 & {D0580101} \\
D05802 & {D0580201} \\
D11701 & {D1170101} {D1170102} {D1170103} \\
D12001 & {D1200101} \\
D12003 & {D1200301} \\
D12006 & {D1200601} {D1201601} \\
D12008 & {D1200801} \\
D12017 & {D1201701} \\
D15801 & {D1580101} {D1580102} \\
D90301 & {D9030101} \\
D90302 & {D9030201} {D9030202} \\
D90305 & {D9030501} \\
E12101 & {E1210101} {E1210201} \\
I20509 & {I2050901} \\
M10103 & {M1010301} \\
M10704 & {M1070414} {M1070415} {M1070420} {M1070421} {M1070423} {M1070424} {M1070427} {S4050201} {S4050202} {S4050204} \\
P10409 & {P1040901} \\
P10411 & {P1041101} {P1041102} {P1041103} \\
P10414 & {P1041403} \\
P10418 & {P1041801} {S5053701} \\
P10420 & {P1042003} {S4050903} {S4050904} \\
P10421 & {P1042101} {P1042105} {S4050501} {S4050502} {S4050503} {S4050504} {S5130201} \\
P10425 & {P1042501} {P1042601} \\
P10429 & {P1042901} {P1042902} {S4057901} {S4057902} {S4057903} {S4057904} {S4057905} {S5054401} \\
P11003 & {P1100301} {P1100302} \\
P11607 & {P1160701} \\
P12011 & {P1201112} \\
P12012 & {P1201213} \\
P19802 & {P1980202} \\
P20406 & {P2040601} \\
P20408 & {P2040802} {P2040803} {S4050801} \\
P20410 & {P2041002} {S4053601} \\
P20411 & {P2041102} {P2041103} {P2041104} \\
P20419 & {P2041901} \\
P20421 & {P2042101} \\
P20422 & {P2042201} {P2042202} {P2042203} {S4055602} {S4055603} {S4055607} {S4055608} {S4055609} \\
P20423 & {P2042301} \\
P20516 & {P2051602} {P2051603} {P2051604} {P2051605} {P2051606} {P2051607} {P2051608} {P2051609} {S4055701} {S4055702} \\
& {S4055703} {S4055704} {S4055705} {S4055707} \\
P20517 & {P2051701} {P2051702} {P2051703} \\
P30208 & {P3020802} \\
P30314 & {P3031401} {P3031402} \\
Q10803 & {Q1080303} \\
Q11002 & {Q1100201} \\
S10102 & {S1010206} {S1010207} {S4056301} {S4056302} {S4056303} {S4056304} {S4056305} {S4056306} {S4056307} \\
S30402 & {S3040203} {S3040204} {S3040205} {S3040206} {S3040207} {S3040208} {S4059601} {S4059602} \\
S40501 & {S4050101} {S4050103} \\
S40504 & {S4050401} \\
S40506 & {S4050601} \\
S40507 & {S4050701} \\
S40510 & {S4051001} {S4051002} \\
S40512 & {S4051201} \\
S40513 & {S4051301} \\
S40514 & {S4051401} {S4051402} \\
S40515 & {S4051501} {S4051502} \\
S40518 & {S4051802} {S4051803} {S4051804} {S4051805} \\
S40521 & {S4052101} \\
S40522 & {S4052201} \\
S40523 & {S4052301} {S4052302} \\
S40524 & {S4052402} {S4052403} {S4052404} {S4052405} {S4052406} \\
S40525 & {S4052501} \\
S40526 & {S4052601} \\
S40528 & {S4052801} {S4052802} \\
S40529 & {S4052901} {S4052902} \\
S40531 & {S4053101} \\
S40532 & {S4053201} {S4053202} {S5053201} \\
S40533 & {S4053301} {S4053302} \\
S40534 & {S4053401} \\
S40535 & {S4053501} \\
S40537 & {S4053701} {S4053702} \\
S40540 & {S4054001} {S4054002} {S4054003} {S4054004} {S4054005} \\
S40541 & {S4054101} \\
S40542 & {S4054201} \\
S40543 & {S4054301} \\
S40548 & {S4054801} \\
S40549 & {S4054901} {S4054902} \\
S40550 & {S4055001} {S4058001} \\
S40552 & {S4055201} \\
S40553 & {S4055301} \\
S40555 & {S4055501} {S4055502} {S4055503} {S4055504} \\
S40558 & {S4055801} \\
S40559 & {S4055901} \\
S40560 & {S4056001} {S4056002} {S4058401} \\
S40561 & {S4056101} \\
S40562 & {S4056201} \\
S40564 & {S4056401} \\
S40565 & {S4056501} {S4056502} \\
S40566 & {S4056601} {S4056602} {S4056603} {S4056604} \\
S40568 & {S4056801} {S4056802} \\
S40570 & {S4057001} {S4057002} \\
S40572 & {S4057201} \\
S40573 & {S4057301} {S4057302} {S4057303} {S5220101} \\
S40574 & {S4057401} \\
S40577 & {S4057701} \\
S40581 & {S4058101} {S4058102} \\
S40582 & {S4058201} {S4058202} {S4058203} {S5054101} \\
S40583 & {S4058301} \\
S40587 & {S4058701} \\
S40588 & {S4058801} \\
S40589 & {S4058901} \\
S40590 & {S4059001} \\
S40594 & {S4059401} \\
S50502 & {S4057601} {S4057602} {S4057603} {S4057604} {S5050201} {S5050202} {S5230601} \\
S50504 & {S5050402} {S5050403} \\
S50505 & {S5050501} \\
S50506 & {S5050601} \\
S50507 & {S5050701} \\
S50508 & {S5050801} \\
S50509 & {S5050901} {S5050902} \\
S50510 & {S5051001} {S5051101} \\
S50512 & {S5051201} \\
S50514 & {S5051401} \\
S50515 & {S5051501} \\
S50516 & {S5051601} \\
S50517 & {S5051701} \\
S50518 & {S5051801} {S5052101} \\
S50522 & {S5052201} \\
S50523 & {S5052301} \\
S50525 & {S5052501} \\
S50527 & {S5052701} \\
S50529 & {S5052901} \\
S50530 & {S5053001} \\
S50533 & {S5053301} \\
S50536 & {S5053601} \\
S50538 & {S5053801} \\
S50539 & {S5053901} \\
S50540 & {S5054001} \\
S50542 & {S5054201} \\
S50543 & {S5054301} \\
S51303 & {S5130301} \\
S51401 & {S5140102} \\
S51601 & {S4051701} {S4059101} {S5160101} \\
S52001 & {S5200101} \\
S52301 & {S5050101} {S5050102} {S5050103} {S5230101} {S5230102} {S5230201} {S5230202} {S5230301} {S5230302} {S5230401} \\
& {S5230402} {S5230501} {S5230502} {S5230503} {S5230504} {S5230505} {S5230506} {S5230507} {S5230508} \\
S52307 & {S5051301} {S5051302} {S5230703} {S5230704} \\
S52309 & {S5230901} \\
Z90702 & {Z9070201} \\
Z90706 & {Z9070601} \\
Z90708 & {Z9070801} \\
Z90709 & {Z9070901} \\
Z90711 & {Z9071101} \\
Z90712 & {Z9071201} \\
Z90714 & {Z9071401} \\
Z90715 & {Z9071501} \\
Z90719 & {Z9071901} \\
Z90721 & {Z9072101} \\
Z90722 & {Z9072201} \\
Z90725 & {Z9072501} \\
Z90726 & {Z9072601} {Z9072602} \\
Z90727 & {Z9072701} \\
Z90733 & {Z9073301} \\
Z90735 & {Z9073501} \\
Z90736 & {Z9073601} \\
Z90737 & {Z9073701} {Z9073702}
\enddata
\end{deluxetable}

\clearpage
\LongTables
\tabletypesize{\scriptsize}
\begin{landscape}
\begin{deluxetable}{llrrrcrcrrcrc} 
\tablecolumns{13} 
\tablewidth{0pt}
\tablecaption{\label{tab_detections} \ion{O}{6} $\lambda 1032$ Detections}
\tablehead{ 
\colhead{Sight}  & \colhead{Target} & \colhead{$l$}   & \colhead{$b$}   & \colhead{$t$\tablenotemark{b}} & & \colhead{Intensity\tablenotemark{d}} & \colhead{Wavelength\tablenotemark{e}} &
\colhead{FWHM\tablenotemark{f}}   & \colhead{$v_{\rm LSR}$} & \colhead{$E(B-V)$\tablenotemark{g}} &  \colhead{SXR\tablenotemark{h}} & \colhead{H$_{\alpha}$\tablenotemark{i}} \\
\colhead{Line}    & \colhead{Name\tablenotemark{a}} & \colhead{(deg)} & \colhead{(deg)} & \colhead{(s)} & \colhead{Bin\tablenotemark{c}} & \colhead{($10^3$ LU)}   & \colhead{(\AA)} &
\colhead{(km s$^{-1}$)} & \colhead{(km s$^{-1}$)}   & \colhead{(mag)} & \colhead{(RU)} & \colhead{($10^3$ LU)} }
\startdata 
\cutinhead{$3 \sigma$ Detections} 
A04604 & NGC5253               & 314.9 & $ 30.1$ & 18252 & 14 & $4.5 \pm 1.4$ & $1031.93 \pm 0.08$ & $150 \pm 70$ & $   2 \pm 22$ & 0.056 & $ 820 \pm 40$ & $ 134 \pm 3$ \\
A11703 & Virgo                 & 284.0 & $ 74.5$ & 10931 & \phn8 & $3.7 \pm 1.2$ & $1031.95 \pm 0.06$ & $ 60 \pm 20$ & $   9 \pm 18$ & 0.023 & $4520 \pm 90$ & \nodata \\
B12901 & SKY-033255-632751     & 278.6 & $-45.3$ & 83579 & \phn8 & $2.0 \pm 0.6$ & $1032.64 \pm 0.06$ & $110 \pm 70$ & $ 192 \pm 19$ & 0.160 & $ 590 \pm 40$ & \nodata \\
C06401 & NGC5846               &   0.4 & $ 48.8$ & 39127 & \phn8 & $2.9 \pm 0.8$ & $1031.46 \pm 0.05$ & $ 40 \pm 40$ & $-122 \pm 13$ & 0.056 & $ 740 \pm 50$ & $  65 \pm 3$ \\
C07601 & NGC6752               & 336.5 & $-25.6$ & 56225 & \phn8 & $1.8 \pm 0.4$ & $1031.87 \pm 0.02$ & $  6 \pm 10$ & $ -17 \pm \phn6$ & 0.056 & $ 490 \pm 80$ & \nodata \\
C10503 & MS2318.2-4220         & 348.1 & $-66.3$ & 97954 & \phn8 & $1.8 \pm 0.4$ & $1031.92 \pm 0.05$ & $ 90 \pm 40$ & $  -5 \pm 13$ & 0.021 & $ 880 \pm 70$ & \nodata \\
C16501 & HD22049               & 195.8 & $-48.1$ & 85441 & \phn8 & $2.1 \pm 0.5$ & $1031.97 \pm 0.04$ & $ 50 \pm 50$ & $  -2 \pm 11$ & 0.039 & $1470 \pm 60$ & $ 476$ \\
D04001 & NGC625                & 273.7 & $-73.1$ & 43428 & \phn8 & $2.6 \pm 0.8$ & $1032.03 \pm 0.06$ & $ 90 \pm 50$ & $  19 \pm 18$ & 0.016 & $1040 \pm 40$ & \nodata \\
D90305 & Fairall 917           & 346.6 & $-39.5$ & 21796 & 14 & $5.5 \pm 1.5$ & $1031.96 \pm 0.08$ & $180 \pm 50$ & $  11 \pm 24$ & 0.035 & $ 790 \pm 50$ & \nodata \\
I20509 & Background            & 315.0 & $-41.3$ & 93437 & \phn8 & $3.3 \pm 0.6$ & $1032.05 \pm 0.05$ & $170 \pm 40$ & $  26 \pm 16$ & 0.036 & $ 890 \pm 60$ & \nodata \\
P10411 & WD0455-282            & 229.3 & $-36.2$ & 46240 & \phn8 & $4.3 \pm 0.9$ & $1031.95 \pm 0.03$ & $ 60 \pm 40$ & $ -12 \pm \phn9$ & 0.023 & $ 900 \pm 40$ & $ 211 \pm 3$ \\
P10414 & HD039659              & 166.2 & $ 10.5$ &  7969 & \phn8 & $3.2 \pm 1.0$ & $1031.14 \pm 0.02$ & $  3 \pm 20$ & $-231 \pm  7$ & 0.233 & $ 320 \pm 30$ & $ 365 \pm 3$ \\
P10429 & WD1631+781            & 111.3 & $ 33.6$ & 90149 & \phn8 & $2.9 \pm 0.8$ & $1031.71 \pm 0.14$ & $300 \pm 50$ & $ -50 \pm 40$ & 0.039 & $ 583 \pm 22$ & $  56 \pm 3$ \\
P12011 & JUP-DTOH1             & 194.3 & $ 13.2$ &  4731 & 14 & $8.8 \pm 2.4$ & $1031.94 \pm 0.10$ & $200 \pm 40$ & $  -8 \pm 28$ & 0.056 & $1030 \pm 40$ & $ 352 \pm 3$ \\
P20411 & WD0004+330            & 112.5 & $-28.7$ & 48131 & \phn8 & $2.1 \pm 0.6$ & $1031.90 \pm 0.05$ & $ 60 \pm 40$ & $  -1 \pm 15$ & 0.049 & $ 290 \pm 40$ & $ 115 \pm 3$ \\
P20422 & WD2004-605            & 336.6 & $-32.9$ & 76972 & \phn8 & $4.8 \pm 0.6$ & $1031.91 \pm 0.03$ & $130 \pm 30$ & $  -7 \pm \phn8$ & 0.049 & $ 880 \pm 80$ & \nodata \\
P20516 & LSE 44                & 313.4 & $ 13.5$ & 85811 & \phn8 & $3.5 \pm 0.6$ & $1031.83 \pm 0.04$ & $100 \pm 20$ & $ -28 \pm 12$ & 0.124 & $ 490 \pm 50$ & \nodata \\
S30402 & HD224151              & 115.4 & $ -4.6$ & 52925 & \phn8 & $2.9 \pm 0.9$ & $1032.65 \pm 0.12$ & $210 \pm 80$ & $ 217 \pm 34$ & 0.691 & $ 325 \pm 22$ & $1054$ \\
S40506 & HD093840-BKGD         & 282.1 & $ 11.1$ & 12962 & \phn8 & $4.0 \pm 1.3$ & $1031.91 \pm 0.05$ & $ 20 \pm 100$ & $ -14 \pm 13$ & 0.167 & $ 650 \pm 40$ & \nodata \\
S40543 & NCVZ-BKGD             & 100.8 & $ 44.8$ & 13915 & \phn8 & $3.1 \pm 1.0$ & $1032.65 \pm 0.05$ & $ 30 \pm 20$ & $ 224 \pm 16$ & 0.031 & $ 960 \pm 30$ & $  31$ \\
S40548 & CVZ-BKGD              &  95.4 & $ 36.1$ & 98216 & \phn8 & $2.8 \pm 0.6$ & $1031.82 \pm 0.05$ & $150 \pm 40$ & $ -15 \pm 15$ & 0.024 & $ 939 \pm 14$ & $  81$ \\
S40549 & NCVZ-BKGD             & 107.0 & $ 48.8$ &  2406 & 14 & $8.2 \pm 2.3$ & $1031.29 \pm 0.03$ & $  2 \pm 20$ & $-172 \pm \phn9$ & 0.021 & $1121 \pm 23$ & $  48 \pm 3$ \\
S40560 & WD1725+586-BKGD       &  87.2 & $ 33.8$ & 22975 & \phn8 & $4.2 \pm 1.4$ & $1031.82 \pm 0.06$ & $110 \pm 90$ & $ -13 \pm 16$ & 0.037 & $ 756 \pm 19$ & $ 106 \pm 4$ \\
S40590 & HE2-138-BKGD          & 319.7 & $ -9.4$ & 10816 & \phn8 & $3.3 \pm 0.9$ & $1031.77 \pm 0.02$ & $  3 \pm 10$ & $ -49 \pm \phn5$ & 0.109 & $ 410 \pm 40$ & \nodata \\
S50504 & HD003827-BKGD         & 120.8 & $-23.2$ & 14137 & \phn8 & $4.4 \pm 0.9$ & $1031.86 \pm 0.02$ & $  3 \pm 40$ & $ -15 \pm \phn7$ & 0.063 & $ 560 \pm 30$ & $ 224$ \\
S50508 & LS982-BKGD            & 257.1 & $ -3.9$ & 10515 & \phn8 & $9.1 \pm 1.8$ & $1031.97 \pm 0.03$ & $ 30 \pm 30$ & $  -4 \pm \phn8$ & 1.260 & $ 670 \pm 30$ & \nodata \\
S50509 & HD26976-BKGD          & 200.8 & $-38.0$ & 16931 & 14 & $4.9 \pm 1.4$ & $1031.84 \pm 0.09$ & $190 \pm 50$ & $ -41 \pm 27$ & 0.096 & $1160 \pm 50$ & $ 364$ \\
S50510 & WD0232+035-BKGD       & 166.0 & $-50.3$ & 34523 & \phn8 & $2.2 \pm 0.6$ & $1031.80 \pm 0.03$ & $  6 \pm 30$ & $ -45 \pm \phn8$ & 0.043 & $ 570 \pm 40$ & $  61 \pm 3$ \\
Z90715 & 2MASXiJ1622346+735943 & 106.8 & $ 35.3$ & 14596 & 14 & $4.4 \pm 1.4$ & $1031.92 \pm 0.09$ & $160 \pm 70$ & $  13 \pm 27$ & 0.031 & $ 685 \pm 20$ & $  78 \pm 3$ \\
\cutinhead{$2 \sigma$ Detections} 
A09401 & Mrk153                & 156.7 & $ 56.0$ & 20589 & 14 & $3.2 \pm 1.3$ & $1031.86 \pm 0.10$ & $140 \pm 100$ & $ -14 \pm 30$ & 0.013 & $1000 \pm 30$ & $  22 \pm 3$ \\
A10001 & ALPHA-TRA             & 321.5 & $-15.3$ &  9409 & 14 & $2.7 \pm 1.3$ & $1031.82 \pm 0.13$ & $100 \pm \phn20$ & $ -35 \pm 38$ & 0.105 & $ 500 \pm 60$ & \nodata \\
B00302 & REJ-1043+490          & 162.7 & $ 57.0$ & 28618 & 14 & $2.6 \pm 1.0$ & $1031.92 \pm 0.17$ & $230 \pm \phn80$ & $   2 \pm 49$ & 0.013 & $1110 \pm 40$ & $  15 \pm 3$ \\
B00303 & REJ-1059+514          & 156.3 & $ 57.8$ & 24823 & \phn8 & $2.9 \pm 1.0$ & $1031.70 \pm 0.14$ & $200 \pm \phn40$ & $ -60 \pm 42$ & 0.011 & $1140 \pm 40$ & \nodata \\
B01802 & NGC604                & 133.8 & $-31.2$ &  5940 & 14 & $7.4 \pm 4.9$ & $1032.53 \pm 0.32$ & $270 \pm 230$ & $ 176 \pm 93$ & 0.046 & $ 440 \pm 30$ & $ 130 \pm 3$ \\
C10101 & NGC6543-N-extension   &  96.5 & $ 30.0$ & 35164 & \phn8 & $3.1 \pm 1.5$ & $1031.71 \pm 0.19$ & $260 \pm 140$ & $ -46 \pm 55$ & 0.045 & $ 529 \pm  4$ & $1588 \pm 4$ \\
D05802 & WD1528+487            &  78.9 & $ 52.7$ & 12136 & 14 & $6.8 \pm 2.5$ & $1031.99 \pm 0.15$ & $200 \pm \phn30$ & $  33 \pm 44$ & 0.013 & $1260 \pm 40$ & $  25 \pm 3$ \\
D12001 & IC289                 & 138.8 & $  2.8$ &  5464 & \phn8 & $5.4 \pm 2.5$ & $1031.90 \pm 0.10$ & $ 80 \pm \phn60$ & $  -5 \pm 30$ & 1.208 & $ 410 \pm 30$ & $ 836 \pm 4$ \\
D12006 & NGC6826-POS2          &  83.6 & $ 12.8$ &  5108 & 14 & $3.0 \pm 2.0$ & $1032.19 \pm 0.11$ & $ 20 \pm 210$ & $  92 \pm 33$ & 0.111 & $ 515 \pm 18$ & $1470 \pm 4$ \\
D12008 & NGC7354               & 107.8 & $  2.3$ &  3009 & 14 & $6.6 \pm 3.1$ & $1033.00 \pm 0.14$ & $100 \pm \phn30$ & $ 323 \pm 41$ & 2.207 & $ 364 \pm 23$ & $1432 \pm 4$ \\
E12101 & R-Aqr-jet-NE          &  66.5 & $-70.3$ & 20959 & 14 & $4.4 \pm 2.3$ & $1033.58 \pm 0.32$ & $450 \pm 180$ & $ 478 \pm 92$ & 0.025 & $ 590 \pm 40$ & \nodata \\
P10418 & HD61421               & 213.7 & $ 13.0$ & 13934 & 14 & $4.5 \pm 1.8$ & $1031.84 \pm 0.08$ & $130 \pm \phn90$ & $ -40 \pm 24$ & 0.061 & $ 600 \pm 40$ & $ 240$ \\
P10420 & WD1034+001            & 247.6 & $ 47.8$ & 14964 & \phn8 & $3.0 \pm 1.3$ & $1032.50 \pm 0.07$ & $ 30 \pm \phn30$ & $ 159 \pm 19$ & 0.075 & $ 550 \pm 30$ & $ 228 \pm 4$ \\
P11003 & HS1307+4617           & 113.0 & $ 70.7$ & 43176 & \phn8 & $1.3 \pm 0.5$ & $1031.71 \pm 0.10$ & $100 \pm \phn40$ & $ -51 \pm 30$ & 0.009 & $ 990 \pm 40$ & $  37 \pm 2$ \\
P20408 & WD1211+332            & 175.0 & $ 80.0$ & 19616 & 14 & $3.2 \pm 1.5$ & $1031.92 \pm 0.11$ & $140 \pm 100$ & $   5 \pm 31$ & 0.012 & $1040 \pm 40$ & $  35$ \\
P20410 & WD1800+685            &  98.7 & $ 29.8$ & 19737 & \phn8 & $2.7 \pm 1.1$ & $1031.84 \pm 0.08$ & $ 80 \pm \phn60$ & $ -10 \pm 24$ & 0.056 & $ 613 \pm \phn7$ & $ 115 \pm 3$ \\
P20423 & WD2156-546            & 339.7 & $-48.1$ &  4403 & 14 & $6.6 \pm 4.5$ & $1032.61 \pm 0.31$ & $250 \pm 220$ & $ 196 \pm 91$ & 0.025 & $2020 \pm 60$ & \nodata \\
P20517 & LS 1274               & 277.0 & $ -5.3$ & 35945 & \phn8 & $4.9 \pm 1.7$ & $1031.81 \pm 0.15$ & $290 \pm 120$ & $ -47 \pm 45$ & 0.550 & $ 400 \pm 30$ & \nodata \\
P30314 & RE\_J1738+66          &  96.9 & $ 32.0$ & 21810 & \phn8 & $2.5 \pm 0.9$ & $1031.84 \pm 0.06$ & $ 30 \pm \phn30$ & $  -8 \pm 17$ & 0.044 & $ 614 \pm \phn7$ & $ 128 \pm 3$ \\
Q11002 & HD125162              &  87.0 & $ 64.7$ &  8203 & 14 & $8.2 \pm 3.3$ & $1031.81 \pm 0.31$ & $450 \pm 110$ & $ -21 \pm 91$ & 0.010 & $1270 \pm 30$ & $  33$ \\
S40504 & NGC2392-BKGD          & 197.9 & $ 17.4$ &  1639 & 14 & $7.8 \pm 4.0$ & $1030.63 \pm 0.10$ & $ 90 \pm 110$ & $-387 \pm 30$ & 0.051 & $1680 \pm 50$ & $ 397 \pm 3$ \\
S40507 & HD96548-BKG           & 292.3 & $ -4.8$ & 13133 & \phn8 & $2.9 \pm 1.2$ & $1031.65 \pm 0.03$ & $  1 \pm 350$ & $ -90 \pm 10$ & 0.810 & $ 303 \pm 24$ & \nodata \\
S40510 & NCVZ-BKGD             & 117.2 & $ 46.3$ & 40524 & \phn8 & $1.0 \pm 0.5$ & $1030.46 \pm 0.04$ & $  5 \pm \phn50$ & $-415 \pm 13$ & 0.012 & $1140 \pm 30$ & $   9 \pm 4$ \\
S40513 & PG1032+406-BKGD       & 178.9 & $ 59.0$ & 12206 & 14 & $2.9 \pm 1.9$ & $1031.79 \pm 0.11$ & $120 \pm 170$ & $ -39 \pm 32$ & 0.013 & $1500 \pm 50$ & $  51$ \\
S40514 & HD163522-BKGD         & 349.6 & $ -9.1$ & 21826 & \phn8 & $6.3 \pm 2.8$ & $1032.27 \pm 0.29$ & $410 \pm 190$ & $ 104 \pm 85$ & 0.182 & $ 260 \pm 60$ & \nodata \\
S40529 & HD013268-BKGD         & 134.0 & $ -5.0$ & 11803 & \phn8 & $2.1 \pm 1.1$ & $1031.42 \pm 0.10$ & $ 60 \pm \phn30$ & $-142 \pm 29$ & 0.411 & $ 373 \pm 23$ & $ 443 \pm 3$ \\
S40534 & EG50-BKGD             & 178.3 & $ 15.4$ &  2204 & 14 & $9.5 \pm 3.7$ & $1031.53 \pm 0.06$ & $ 40 \pm 120$ & $-122 \pm 18$ & 0.129 & $ 400 \pm 30$ & $ 102 \pm 3$ \\
S40537 & RE0503-289-BKGD       & 230.7 & $-34.9$ & 20940 & \phn8 & $2.7 \pm 1.1$ & $1032.04 \pm 0.11$ & $ 60 \pm \phn90$ & $  13 \pm 33$ & 0.015 & $1220 \pm 40$ & $ 249 \pm 4$ \\
S40541 & SWUMa-BKGD            & 164.8 & $ 37.0$ &  2285 & 14 & $9.2 \pm 4.9$ & $1032.05 \pm 0.30$ & $300 \pm 160$ & $  36 \pm 87$ & 0.037 & $ 610 \pm 50$ & $  54$ \\
S40552 & NCVZ-BKGD             & 105.3 & $ 54.3$ &  2063 & 14 & $6.8 \pm 5.7$ & $1030.78 \pm 0.13$ & $ 70 \pm 200$ & $-319 \pm 38$ & 0.012 & $1310 \pm 40$ & $  24 \pm 2$ \\
S40555 & PG1520+525-BKGD       &  85.4 & $ 52.3$ & 31588 & \phn8 & $2.1 \pm 1.0$ & $1031.26 \pm 0.08$ & $ 80 \pm 100$ & $-177 \pm 22$ & 0.015 & $1240 \pm 30$ & $  35$ \\
S40568 & V3885-Sgr-BKGD        & 357.5 & $-27.8$ & 11873 & \phn8 & $4.6 \pm 1.8$ & $1031.91 \pm 0.08$ & $100 \pm \phn80$ & $   0 \pm 22$ & 0.062 & $ 690 \pm 50$ & \nodata \\
S40581 & HD113001-BKGD         & 110.3 & $ 81.7$ & 14244 & 14 & $3.6 \pm 1.4$ & $1031.79 \pm 0.14$ & $200 \pm \phn40$ & $ -29 \pm 42$ & 0.010 & $1090 \pm 30$ & $  26 \pm 3$ \\
S40594 & HD219188-BKGD         &  83.0 & $-50.2$ &  9218 & 14 & $3.2 \pm 1.6$ & $1034.03 \pm 0.12$ & $100 \pm \phn60$ & $ 614 \pm 36$ & 0.072 & $ 440 \pm 30$ & $  51 \pm 3$ \\
S50505 & NEAR-WD2211-495-BKGD  & 345.6 & $-52.2$ & 11026 & 14 & $2.7 \pm 1.3$ & $1031.92 \pm 0.11$ & $110 \pm \phn90$ & $  -4 \pm 33$ & 0.016 & $1010 \pm 40$ & \nodata \\
S50515 & MRK876-BKGD           &  90.5 & $ 56.6$ &  5832 & \phn8 & $2.9 \pm 1.4$ & $1032.75 \pm 0.03$ & $  3 \pm \phn20$ & $ 255 \pm  7$ & 0.016 & $1080 \pm 30$ & $  62 \pm 3$ \\
S50523 & EC11481-2303-BKGD     & 285.3 & $ 37.4$ &  5042 & 14 & $5.6 \pm 3.3$ & $1031.60 \pm 0.40$ & $300 \pm \phn40$ & $ -99 \pm 117$ & 0.056 & $ 660 \pm 30$ & $ 112 \pm 3$ \\
S50529 & PG0242+132-BKGD       & 160.6 & $-41.0$ &  3572 & 14 & $4.9 \pm 2.0$ & $1030.41 \pm 0.07$ & $  5 \pm \phn70$ & $-448 \pm 20$ & 0.104 & $ 510 \pm 40$ & $  83 \pm 3$ \\
S50540 & BD+523210-BKGD        & 102.4 & $ -3.4$ & 10035 & \phn8 & $2.6 \pm 1.2$ & $1031.18 \pm 0.03$ & $  3 \pm \phn20$ & $-203 \pm  8$ & 0.460 & $ 317 \pm 23$ & $1450 \pm 4$ \\
S50542 & PG0919+272-BKGD       & 200.5 & $ 43.9$ &  5307 & 14 & $5.8 \pm 2.5$ & $1032.03 \pm 0.12$ & $160 \pm 100$ & $  23 \pm 36$ & 0.023 & $ 690 \pm 30$ & $  59 \pm 3$ \\
S51601 & HD104994-BKGD         & 297.6 & $  0.3$ & 17129 & \phn8 & $7.9 \pm 3.4$ & $1032.04 \pm 0.26$ & $410 \pm 150$ & $  25 \pm 75$ & 3.504 & $ 350 \pm 40$ & \nodata \\
Z90711 & RXJ1729.1+7033        & 101.3 & $ 32.3$ & 21722 & \phn8 & $1.6 \pm 0.6$ & $1031.74 \pm 0.04$ & $  5 \pm 100$ & $ -39 \pm 13$ & 0.034 & $ 691 \pm 13$ & $  83 \pm 3$ \\
Z90726 & IRAS05595-5756        & 266.5 & $-29.4$ & 21079 & 14 & $2.0 \pm 0.8$ & $1032.10 \pm 0.09$ & $100 \pm \phn30$ & $  32 \pm 27$ & 0.045 & $ 511 \pm 17$ & \nodata \\
Z90733 & NGC3735               & 131.7 & $ 45.3$ &  6967 & 14 & $3.5 \pm 1.4$ & $1031.41 \pm 0.06$ & $ 20 \pm \phn90$ & $-140 \pm 18$ & 0.017 & $1030 \pm 30$ & $  24 \pm 3$
\enddata 
\tablenotetext{a}{Target names are taken from the data file headers.  Some have been modified for this table.}
\tablenotetext{b}{Exposure time is night only.}
\tablenotetext{c}{Spectral binning in 0.013 \AA\ pixels.}
\tablenotetext{d}{1\,LU = 1~photon~s$^{-1}$\,cm$^{-2}$\,sr$^{-1}$.}
\tablenotetext{e}{Wavelengths are heliocentric.}
\tablenotetext{f}{Gaussian FWHM values include the smoothing imparted by the instrument optics. Values less than $\sim$ 25 \kms\ indicate that the emission does not fill the LWRS aperture.}
\tablenotetext{g}{Extinction from \citealt{Schlegel:98}.}
\tablenotetext{h}{{\em ROSAT} 1/4 keV emission from \citealt{Snowden:97}.
1 RU = $10^{-6}$~counts~s$^{-1}$\,arcmin$^{-2}$.}
\tablenotetext{i}{H$\alpha$ intensity integrated over the velocity range $-80$ to +80 \kms\ from \citealt{Haffner:03}.
Values without error bars were derived from the average of the surrounding pointings.  Data are available only for declinations above $-30$\degr.}
\end{deluxetable} 
\clearpage
\end{landscape}

\clearpage
\LongTables

\tabletypesize{\scriptsize}
\begin{deluxetable}{clrrrccrc} 
\tablecolumns{9} 
\tablewidth{0pt}
\tablecaption{\label{tab_limits} O VI $\lambda 1032$ $3\sigma$ Upper Limits}
\tablehead{ 
\colhead{Sight}  & \colhead{Target} & \colhead{$l$}   & \colhead{$b$}   & \colhead{$t$\tablenotemark{b}}    & \colhead{$3\sigma$ Limit\tablenotemark{c}} &
\colhead{$E(B-V)$\tablenotemark{d}} &  \colhead{SXR\tablenotemark{e}} & \colhead{H$_{\alpha}$\tablenotemark{f}} \\
\colhead{Line}    & \colhead{Name\tablenotemark{a}} & \colhead{(deg)} & \colhead{(deg)} & \colhead{(s)} & \colhead{($10^3$ LU)}   &
\colhead{(mag)} & \colhead{(RU)} & \colhead{($10^3$ LU)} }
\startdata 
A01002 & LB9802                & 299.9 & $-30.7$ &  2022 & \phn 5.8 & \phn 0.077 & $ 630 \pm  40$ & \nodata \\
A03406 & WD2218+706            & 110.9 & $ 11.5$ &  1582 & \phn 5.9 & \phn 0.689 & $ 470 \pm  30$ & $  699 \pm  4$ \\
A03407 & RE-J0558-376          & 243.7 & $-26.1$ &  8143 & \phn 3.1 & \phn 0.036 & $ 770 \pm  30$ & \nodata \\
A04802 & HH47A                 & 267.4 & $ -7.5$ &  6297 & \phn 3.9 & \phn 1.095 & $ 570 \pm  24$ & \nodata \\
A05101 & HD200775              & 104.1 & $ 14.2$ & 14786 & \phn 2.0 &     12.290 & $ 364 \pm  19$ & $  361 \pm  4$ \\
A11701 & ComaI                 &  57.6 & $ 88.0$ & 23614 & \phn 1.5 & \phn 0.009 & $4370 \pm  90$ & \nodata \\
A13902 & NGC1068               & 172.1 & $-51.9$ & 77189 & \phn 0.9 & \phn 0.034 & $ 610 \pm  50$ & $   64 \pm  3$ \\
B01805 & NGC5471               & 101.8 & $ 59.6$ &  6303 & \phn 3.5 & \phn 0.010 & $1320 \pm  40$ & $   35 \pm  3$ \\
B04604 & HD206267C             &  99.3 & $  3.7$ & 10072 & \phn 2.6 & \phn 0.963 & $ 283 \pm  17$ & $14237 $ \\
B06801 & GAMMA-CRU             & 300.2 & $  5.6$ & 12104 & \phn 3.0 & \phn 0.789 & $ 390 \pm  30$ & \nodata \\
C02201 & HD6833                & 125.6 & $ -8.0$ & 17498 & \phn 2.2 & \phn 0.444 & $ 440 \pm  50$ & $  602 \pm  4$ \\
C02301 & BETA-GRU              & 346.3 & $-58.0$ &  5677 & \phn 3.0 & \phn 0.009 & $1170 \pm  50$ & \nodata \\
C03701 & PG1626+554            &  84.5 & $ 42.2$ & 29854 & \phn 1.4 & \phn 0.006 & $1430 \pm  30$ & $   38 $ \\
C03702 & NGC7714               &  88.2 & $-55.6$ & 77143 & \phn 0.9 & \phn 0.052 & $ 430 \pm  30$ & $   47 $ \\
C11602 & CD-38-10980           & 341.5 & $  7.3$ &  5939 & \phn 4.0 & \phn 0.885 & $1350 \pm  50$ & \nodata \\
C11603 & HD74389B              & 170.6 & $ 38.6$ & 29341 & \phn 1.6 & \phn 0.030 & $1120 \pm  50$ & $   56 \pm  3$ \\
D05801 & WD2013+400            &  77.0 & $  3.2$ & 30059 & \phn 1.8 & \phn 2.524 & $ 468 \pm  20$ & $13413 \pm 12$ \\
D11701 & 3C249.1               & 130.4 & $ 38.5$ & 39169 & \phn 1.4 & \phn 0.032 & $ 770 \pm  40$ & $   61 \pm  4$ \\
D12003 & IC4593                &  25.3 & $ 40.8$ &  5297 & \phn 4.8 & \phn 0.058 & $ 700 \pm  30$ & $  231 \pm  3$ \\
D12017 & PK342-14D1            & 342.5 & $-14.3$ &  3539 & \phn 4.7 & \phn 0.131 & $ 580 \pm 110$ & \nodata \\
D15801 & WD2020-425            & 358.4 & $-34.5$ & 11638 & \phn 2.3 & \phn 0.054 & $ 880 \pm  50$ & \nodata \\
D90301 & SBS1116+518           & 152.0 & $ 60.1$ & 22775 & \phn 1.7 & \phn 0.014 & $1090 \pm  50$ & $   34 \pm  3$ \\
D90302 & IRAS03335-5626        & 269.4 & $-48.9$ & 33946 & \phn 1.5 & \phn 0.025 & $ 840 \pm  30$ & \nodata \\
M10103 & GD71                  & 192.0 & $ -5.3$ &  9411 & \phn 2.6 & \phn 0.300 & $ 390 \pm  30$ & $  662 \pm  3$ \\
M10704 & WD0439+466            & 158.5 & $  0.5$ & 43550 & \phn 1.7 & \phn 1.139 & $ 310 \pm  30$ & $ 1359 \pm  4$ \\
P10409 & HD029139              & 181.0 & $-20.2$ & 11882 & \phn 2.2 & \phn 0.592 & $ 430 \pm  40$ & $  813 $ \\
P10421 & WD1202+608            & 133.1 & $ 55.7$ & 50097 & \phn 1.2 & \phn 0.015 & $ 920 \pm  40$ & $   19 \pm  2$ \\
P10425 & HD128620/HD128621     & 315.7 & $ -0.7$ & 26651 & \phn 1.6 & \phn 6.505 & $ 340 \pm  30$ & \nodata \\
P11607 & HD38087               & 207.1 & $-16.3$ &  2041 &     11.2 & \phn 0.717 & $ 421 \pm  23$ & $ 5960 \pm  7$ \\
P12012 & JUP-DTOH2             & 194.3 & $ 13.3$ &  4899 & \phn 3.7 & \phn 0.054 & $1070 \pm  40$ & $  352 \pm  3$ \\
P19802 & NGC7293               &  36.2 & $-57.1$ &  4394 & \phn 6.4 & \phn 0.318 & $ 710 \pm  40$ & $ 2163 \pm  5$ \\
P20406 & WD2127-222            &  27.4 & $-43.8$ & 18938 & \phn 2.0 & \phn 0.046 & $ 700 \pm  40$ & $   75 $ \\
P20419 & WD1615-154            & 358.8 & $ 24.2$ & 13891 & \phn 2.5 & \phn 0.305 & $ 690 \pm 100$ & $  812 \pm  4$ \\
P20421 & WD0715-704            & 281.6 & $-23.5$ &  4489 & \phn 4.6 & \phn 0.222 & $ 660 \pm  30$ & \nodata \\
P30208 & LB1566                & 306.4 & $-62.0$ &  8022 & \phn 3.7 & \phn 0.019 & $ 760 \pm  60$ & \nodata \\
Q10803 & PK010-081             &   9.9 & $ -7.6$ &  3989 & \phn 3.5 & \phn 0.326 & $ 280 \pm  30$ & $  895 $ \\
S10102 & WD1634-573            & 329.9 & $ -7.0$ & 67118 & \phn 1.2 & \phn 0.342 & $ 450 \pm  40$ & \nodata \\
S40501 & PG0749+658-BKGD       & 150.5 & $ 31.0$ & 21619 & \phn 1.5 & \phn 0.045 & $ 530 \pm  30$ & $   69 \pm  3$ \\
S40512 & HD163892              &   7.2 & $  0.6$ &  4259 & \phn 4.2 & \phn 7.781 & $ 320 \pm  40$ & $ 2583 \pm  5$ \\
S40515 & HD92809-BKGD          & 286.8 & $ -0.0$ & 14880 & \phn 2.5 & \phn 1.921 & $ 380 \pm  30$ & \nodata \\
S40518 & WD2111+498-BKGD       &  91.4 & $  1.1$ & 19746 & \phn 1.9 & \phn 2.804 & $ 445 \pm  23$ & $ 1552 \pm  4$ \\
S40521 & BD+28D4211-BKGD       &  81.9 & $-19.3$ & 13312 & \phn 2.3 & \phn 0.094 & $ 370 \pm  30$ & $  456 $ \\
S40522 & HD216438-BKGD         & 105.7 & $ -5.1$ &  3945 & \phn 3.7 & \phn 0.399 & $ 280 \pm  30$ & $  897 \pm  4$ \\
S40523 & WOLF1346-BKGD         &  67.2 & $ -9.0$ &  9653 & \phn 2.5 & \phn 0.214 & $ 449 \pm  22$ & $  413 \pm  4$ \\
S40524 & HD203374A-BKGD        & 100.5 & $  8.6$ & 41564 & \phn 1.2 & \phn 1.052 & $ 312 \pm  16$ & $ 2574 \pm  6$ \\
S40525 & A43-BKGD              &  36.1 & $ 17.6$ &  2824 & \phn 4.6 & \phn 0.192 & $ 440 \pm  30$ & $  225 \pm  3$ \\
S40526 & HD156385-BKGD         & 343.2 & $ -4.8$ &  7063 & \phn 3.1 & \phn 0.530 & $ 320 \pm  30$ & \nodata \\
S40528 & HD187459-BKGD         &  68.8 & $  3.9$ & 12190 & \phn 2.3 & \phn 0.677 & $ 390 \pm  25$ & $ 3148 $ \\
S40531 & GD50-BKGD             & 189.0 & $-40.1$ &  5867 & \phn 3.3 & \phn 0.186 & $ 720 \pm  30$ & $ 1402 \pm  4$ \\
S40532 & BD+532820-BKGD        & 101.2 & $ -1.7$ & 20938 & \phn 1.8 & \phn 0.544 & $ 240 \pm  17$ & $ 1518 \pm  5$ \\
S40533 & WZSge-BKGD            &  57.5 & $ -7.9$ &  7651 & \phn 3.6 & \phn 0.315 & $ 560 \pm  30$ & $  424 \pm  4$ \\
S40535 & HD1383-BKGD           & 119.0 & $ -0.9$ &  3684 & \phn 3.9 & \phn 1.539 & $ 410 \pm  40$ & $ 2331 $ \\
S40540 & PG0952+519-BKGD       & 164.1 & $ 49.0$ & 14690 & \phn 2.0 & \phn 0.010 & $1010 \pm  40$ & $   66 \pm  4$ \\
S40542 & Abell78-BKGD          &  81.3 & $-14.9$ &  5480 & \phn 3.6 & \phn 0.164 & $ 280 \pm  30$ & $  389 \pm  4$ \\
S40550 & Z-Cam-BKGD            & 141.4 & $ 32.6$ & 14894 & \phn 2.0 & \phn 0.027 & $ 590 \pm  30$ & $   73 \pm  3$ \\
S40553 & WR42-HD97152-BKGD     & 290.9 & $ -0.5$ & 11844 & \phn 2.9 & \phn 1.571 & $ 300 \pm  30$ & \nodata \\
S40558 & HD102567-BKGD         & 295.6 & $ -0.2$ &  7400 & \phn 3.5 & \phn 3.790 & $ 420 \pm  30$ & \nodata \\
S40559 & BD+354258-BKGD        &  77.2 & $ -4.7$ &  7758 & \phn 3.6 & \phn 0.583 & $ 499 \pm  25$ & $ 2643 $ \\
S40561 & CVZ-BKGD              &  99.3 & $ 43.3$ &  5666 & \phn 3.2 & \phn 0.034 & $ 490 \pm  30$ & $   42 \pm  3$ \\
S40562 & NGC4194-BKGD          & 134.4 & $ 61.8$ & 14414 & \phn 1.9 & \phn 0.016 & $ 860 \pm  40$ & $   38 \pm  2$ \\
S40564 & BD+43D4035-BKGD       & 100.6 & $-13.1$ &  5020 & \phn 3.3 & \phn 0.263 & $ 230 \pm  16$ & $  320 $ \\
S40565 & HD182308-BKGD         &  95.5 & $ 21.3$ &  3630 & \phn 4.0 & \phn 0.103 & $ 416 \pm  15$ & $  235 \pm  3$ \\
S40566 & HD192035-BKGD         &  83.3 & $  7.8$ & 32619 & \phn 1.4 & \phn 0.439 & $ 485 \pm  18$ & $ 1794 \pm  5$ \\
S40570 & BD+48658-BKGD         & 138.1 & $-11.1$ & 23293 & \phn 1.8 & \phn 0.163 & $ 520 \pm  30$ & $  410 \pm  3$ \\
S40572 & HD21291-BKGD          & 141.5 & $  2.9$ & 13773 & \phn 2.2 & \phn 1.317 & $ 290 \pm  22$ & $ 1285 $ \\
S40573 & HD35580-BKGD          & 264.2 & $-34.5$ & 40611 & \phn 1.3 & \phn 0.040 & $ 620 \pm  40$ & \nodata \\
S40574 & PG1051+501-BKGD       & 159.9 & $ 58.5$ &  7500 & \phn 2.8 & \phn 0.019 & $ 930 \pm  30$ & $   40 \pm  3$ \\
S40577 & HDE232522-BKGD        & 130.7 & $ -6.7$ &  5295 & \phn 3.6 & \phn 0.278 & $ 430 \pm  40$ & $  792 \pm  4$ \\
S40582 & CD-61\_1208-BKGD      & 270.1 & $-30.6$ & 42039 & \phn 1.2 & \phn 0.052 & $ 452 \pm  15$ & \nodata \\
S40583 & WD1234+481-BKGD       & 129.8 & $ 69.0$ &  3842 & \phn 4.3 & \phn 0.016 & $1160 \pm  50$ & $   41 \pm  2$ \\
S40587 & BKGD\_1\_mps413       & 161.9 & $ 64.7$ &  8104 & \phn 2.5 & \phn 0.014 & $1220 \pm  40$ & $   27 \pm  3$ \\
S40588 & BKGD\_2\_mps413       & 150.6 & $ 58.9$ & 12295 & \phn 2.2 & \phn 0.017 & $1100 \pm  50$ & $   16 \pm  3$ \\
S40589 & JL\_25-BKGD           & 318.6 & $-29.2$ &  9891 & \phn 2.6 & \phn 0.116 & $ 470 \pm  60$ & \nodata \\
S50502 & G191-B2B-BKGD         & 156.0 & $  7.1$ & 56693 & \phn 1.1 & \phn 0.615 & $ 300 \pm  30$ & $  213 \pm  3$ \\
S50506 & HD201345-BKGD         &  78.4 & $ -9.5$ &  4867 & \phn 4.9 & \phn 0.191 & $ 398 \pm  22$ & $ 1040 \pm  4$ \\
S50507 & HD71634-BKGD          & 273.3 & $-11.5$ & 10265 & \phn 2.8 & \phn 0.261 & $ 400 \pm  30$ & \nodata \\
S50512 & NGC1360-BKGD          & 220.0 & $-54.4$ &  8256 & \phn 2.9 & \phn 0.011 & $2100 \pm 500$ & $  336 \pm  3$ \\
S50514 & HD5679-BKGD           & 123.3 & $ 19.0$ & 10122 & \phn 2.6 & \phn 0.246 & $ 440 \pm  30$ & $  136 \pm  3$ \\
S50516 & HD46223-BKGD          & 206.4 & $ -2.1$ & 13659 & \phn 2.2 & \phn 1.306 & $ 520 \pm  30$ & $ 3601 $ \\
S50517 & HD37367-BKGD          & 179.0 & $ -1.0$ &  9830 & \phn 2.3 & \phn 1.472 & $ 334 \pm  24$ & $  978 $ \\
S50518 & UV0904-02-BKGD        & 233.0 & $ 28.1$ & 12085 & \phn 2.7 & \phn 0.020 & $ 820 \pm  40$ & $   80 \pm  3$ \\
S50522 & HD190864-BKGD         &  72.5 & $  2.0$ &  6213 & \phn 3.5 & \phn 2.175 & $ 397 \pm  22$ & $10786 $ \\
S50525 & PG1519+640-BKGD       & 100.3 & $ 46.2$ &  7467 & \phn 3.4 & \phn 0.016 & $1120 \pm  30$ & $   28 \pm  3$ \\
S50527 & Feige108-BKGD         &  76.8 & $-55.9$ &  4424 & \phn 3.2 & \phn 0.044 & $ 500 \pm  40$ & $   52 \pm  3$ \\
S50530 & DeHt2-BKGD            &  27.7 & $ 16.9$ &  2377 & \phn 6.4 & \phn 0.251 & $ 350 \pm  30$ & $  249 \pm  3$ \\
S50533 & HD29094-BKGD          & 161.8 & $ -4.0$ &  7436 & \phn 2.3 & \phn 0.511 & $ 273 \pm  23$ & $  216 $ \\
S50536 & HD36408-BKGD          & 188.5 & $ -8.9$ & 17691 & \phn 1.7 & \phn 0.579 & $ 306 \pm  22$ & $ 1124 $ \\
S50538 & HD060369-BKGD         & 242.7 & $ -4.3$ &  4838 & \phn 4.1 & \phn 0.513 & $ 700 \pm  40$ & $ 2786 \pm  5$ \\
S50539 & GD1618-BKGD           &  16.2 & $-76.7$ &  3148 & \phn 3.8 & \phn 0.020 & $ 760 \pm  40$ & $   59 \pm  4$ \\
S50543 & PG0101+039-BKGD       & 128.9 & $-58.5$ &  4095 & \phn 3.0 & \phn 0.021 & $ 520 \pm  40$ & $   54 \pm  3$ \\
S51303 & HD092702              & 286.1 & $  1.0$ &  3434 & \phn 5.6 & \phn 1.182 & $ 380 \pm  30$ & \nodata \\
S51401 & CPD-721184            & 299.2 & $-10.9$ &  5955 & \phn 4.0 & \phn 0.241 & $ 220 \pm  30$ & \nodata \\
S52001 & HD186994              &  78.6 & $ 10.1$ &  2271 & \phn 7.9 & \phn 0.201 & $ 337 \pm  19$ & $ 1683 $ \\
S52301 & WD2211-495-BKGD       & 345.8 & $-52.6$ & 58547 & \phn 1.2 & \phn 0.015 & $ 960 \pm  40$ & \nodata \\
S52307 & AGK+81D266-BKGD       & 130.7 & $ 31.9$ & 17424 & \phn 2.5 & \phn 0.026 & $ 780 \pm  30$ & $  108 $ \\
S52309 & HD179406              &  28.2 & $ -8.3$ &  5487 & \phn 3.1 & \phn 0.499 & $ 230 \pm  24$ & $  397 $ \\
Z90702 & CTS0563               & 354.7 & $-49.9$ & 15329 & \phn 1.5 & \phn 0.017 & $ 990 \pm  50$ & \nodata \\
Z90706 & MRK474                &  87.0 & $ 60.6$ &  5759 & \phn 3.3 & \phn 0.033 & $ 770 \pm  30$ & $   55 $ \\
Z90708 & HE2336-5540           & 322.8 & $-58.9$ & 12485 & \phn 3.0 & \phn 0.011 & $1280 \pm  60$ & \nodata \\
Z90709 & PG1246+586            & 123.7 & $ 58.8$ &  8016 & \phn 3.1 & \phn 0.011 & $1150 \pm  40$ & $   30 \pm  2$ \\
Z90712 & FB1229+710            & 125.3 & $ 46.3$ &  5509 & \phn 3.5 & \phn 0.018 & $1000 \pm  30$ & $   60 \pm  3$ \\
Z90714 & 87GB163624.4+713451   & 103.9 & $ 36.2$ & 11021 & \phn 1.9 & \phn 0.045 & $ 554 \pm  14$ & $   88 \pm  3$ \\
Z90719 & ESO116-G18            & 276.2 & $-47.6$ & 10485 & \phn 2.0 & \phn 0.075 & $ 780 \pm  40$ & \nodata \\
Z90721 & PKS0355-483           & 256.2 & $-48.5$ & 11866 & \phn 2.5 & \phn 0.006 & $ 990 \pm  50$ & \nodata \\
Z90722 & NGC1566               & 264.3 & $-43.4$ &  2425 & \phn 4.4 & \phn 0.009 & $ 810 \pm  40$ & \nodata \\
Z90725 & ESO253-G03            & 252.0 & $-33.7$ & 12401 & \phn 2.5 & \phn 0.043 & $ 710 \pm  30$ & \nodata \\
Z90727 & UGC3478               & 151.4 & $ 22.1$ &  2988 & \phn 3.7 & \phn 0.092 & $ 440 \pm  30$ & $  114 \pm  3$ \\
Z90735 & MRK486                &  86.9 & $ 49.4$ & 18011 & \phn 2.0 & \phn 0.014 & $1170 \pm  30$ & $   42 $ \\
Z90736 & NGC6521               &  91.8 & $ 30.2$ & 27742 & \phn 1.3 & \phn 0.041 & $ 722 \pm  10$ & $   61 \pm  3$ \\
Z90737 & FAIRALL333            & 327.7 & $-23.8$ & 29366 & \phn 1.4 & \phn 0.083 & $ 450 \pm  70$ & \nodata \\
\enddata 
\tablenotetext{a}{Target names are taken from the data file headers.}
\tablenotetext{b}{Exposure time is night only.}
\tablenotetext{c}{1\,LU = 1~photon~s$^{-1}$\,cm$^{-2}$\,sr$^{-1}$.}
\tablenotetext{d}{Extinction from \citealt{Schlegel:98}.}
\tablenotetext{e}{{\em ROSAT} 1/4 keV emission from \citealt{Snowden:97}.
1 RU = $10^{-6}$~counts~s$^{-1}$\,arcmin$^{-2}$.}
\tablenotetext{f}{H$\alpha$ intensity integrated over the velocity range $-80$ to +80 \kms\ from \citealt{Haffner:03}.
Values without error bars were derived from the average of
the surrounding pointings.  Data are available only for declinations above $-30$\degr.}
\end{deluxetable} 



\begin{deluxetable}{llcccccc}
\tablecolumns{8}
\tablewidth{0pc} 
\tablecaption{O VI toward Nearby White Dwarfs \label{tab_lehner}} 
\tablehead{
& & & \multicolumn{3}{c}{Absorption} && \colhead{Emission}\\
\cline{4-6} \cline{8-8} 
\colhead{Sight} & \colhead{White} & \colhead{$d$} & \colhead{$\bar{v}_{\rm Helio}$}    &	\colhead{$b$}&   \colhead{$\log N$} && \colhead{$v_{\rm Helio}$}\\
\colhead{Line} & \colhead{Dwarf} & \colhead{(pc)} & \colhead{(\kms)} &   \colhead{(\kms)}&   \colhead{(cm$^{-2}$)} && \colhead{(\kms)}
}
\startdata
P20411 & WD\,0004$+$330 & \phn97 &\phn $-3.8 \pm 3.6 $ &$ 21.3   \pm  4.1 $ & $ 12.79 \pm 0.09$ && \phn $-6 \pm 15$ \\
P10411 & WD\,0455$-$282 & 102  &$-23.6    \pm 4.6 $ &$ 30.1   \pm  7.4 $ & $ 13.42 \pm 0.07$          && \phs \phn $7 \pm \phn9$ \\
P10429 & WD\,1631$+$781& \phn67  &$-16.4    \pm 5.1 $ &  \nodata           & $ 12.52 \pm\,^{0.12}_{0.17} $  && $-64 \pm 40$\\
P20422 & WD\,2004$-$605 & \phn58 &$-23.2    \pm      6.4 $ &  \nodata           & $ 13.00 \pm 0.10 $  && \phn $-6 \pm \phn8$
\enddata
\tablecomments{Stellar distances and absorption data are from \citealt{Savage:06}, who use the interstellar \ion{C}{2} $\lambda 1036.34$ line to tie their velocity scale to that of \citealt{Holberg:98}.  
Emission-line velocities are from this paper.}
\end{deluxetable}


\begin{deluxetable}{lccccccc}
\tablecolumns{8}
\tablewidth{0pt}
\tablecaption{\label{tab_b12901} Observations Included in Sight Line B12901}
\tablehead{
\colhead{Observation}  & \colhead{$l$\tablenotemark{a}}   & \colhead{$b$\tablenotemark{a}}   & \colhead{$t$\tablenotemark{b}}    & \colhead{Intensity\tablenotemark{c}} & \colhead{Wavelength\tablenotemark{d}} &
\colhead{FWHM}        & \colhead{$v_{\rm LSR}$} \\
\colhead{ID}    & \colhead{(deg)} & \colhead{(deg)} & \colhead{(s)} & \colhead{($10^3$ LU)}   & \colhead{(\AA)} &
\colhead{(km s$^{-1}$)} & \colhead{(km s$^{-1}$)} }
\startdata
I2050501, I2050510 & 278.58 & $-45.31$ & 46386 & $3.0 \pm 0.6$ & $1032.69 \pm 0.05$ & $100 \pm 20$ & $206 \pm 13$\\
I2050601 & 278.59 & $-45.30$ & 11120 & $< 2.2$ & \nodata & \nodata & \nodata \\
B1290101, B1290102 & 278.63 & $-45.31$ & 26073 & $< 1.5$ & \nodata & \nodata & \nodata
\enddata
\tablenotetext{a}{Coordinates of LWRS aperture from \citealt{Shelton:03}.}
\tablenotetext{b}{Exposure time is night only.}
\tablenotetext{c}{1\,LU = 1~photon~s$^{-1}$\,cm$^{-2}$\,sr$^{-1}$.}
\tablenotetext{d}{Wavelength is heliocentric.}
\end{deluxetable}


\clearpage

\begin{figure}
\figurenum{2}
\plotone{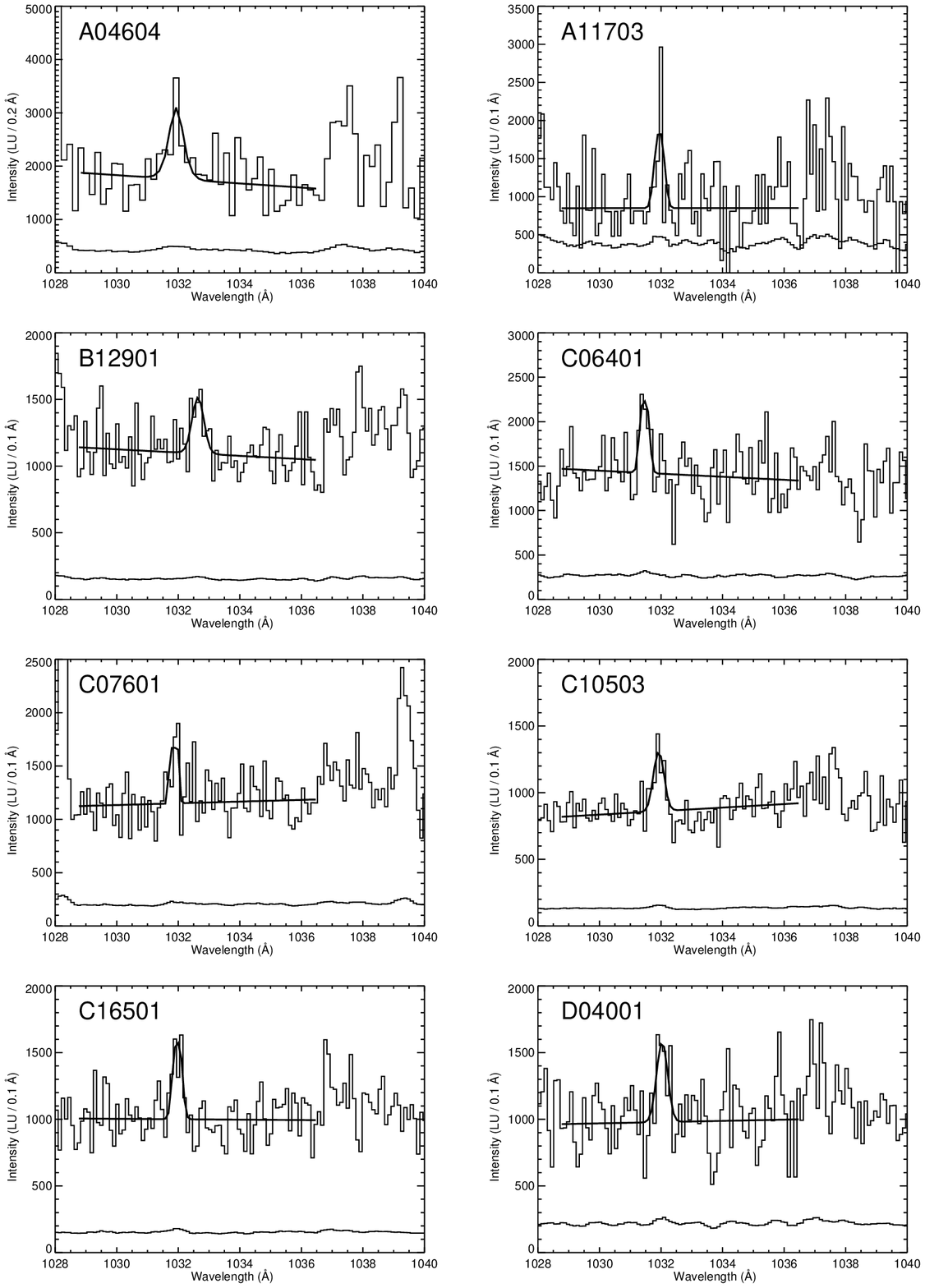}
\caption{High-significance (3\sig) \osix\ $\lambda 1032$ emission features.
The data are binned by the factors given in Table \ref{tab_detections}, and 
best-fit model spectra and error bars are overplotted.
Interstellar \ctwo * $\lambda 1037$ and \osix\ $\lambda 1038$ are present in some spectra,
as are the geocoronal \oone\ $\lambda \lambda 1028, 1039$ lines.}
\label{fig_3sigma}
\end{figure}
\clearpage
%
\plotone{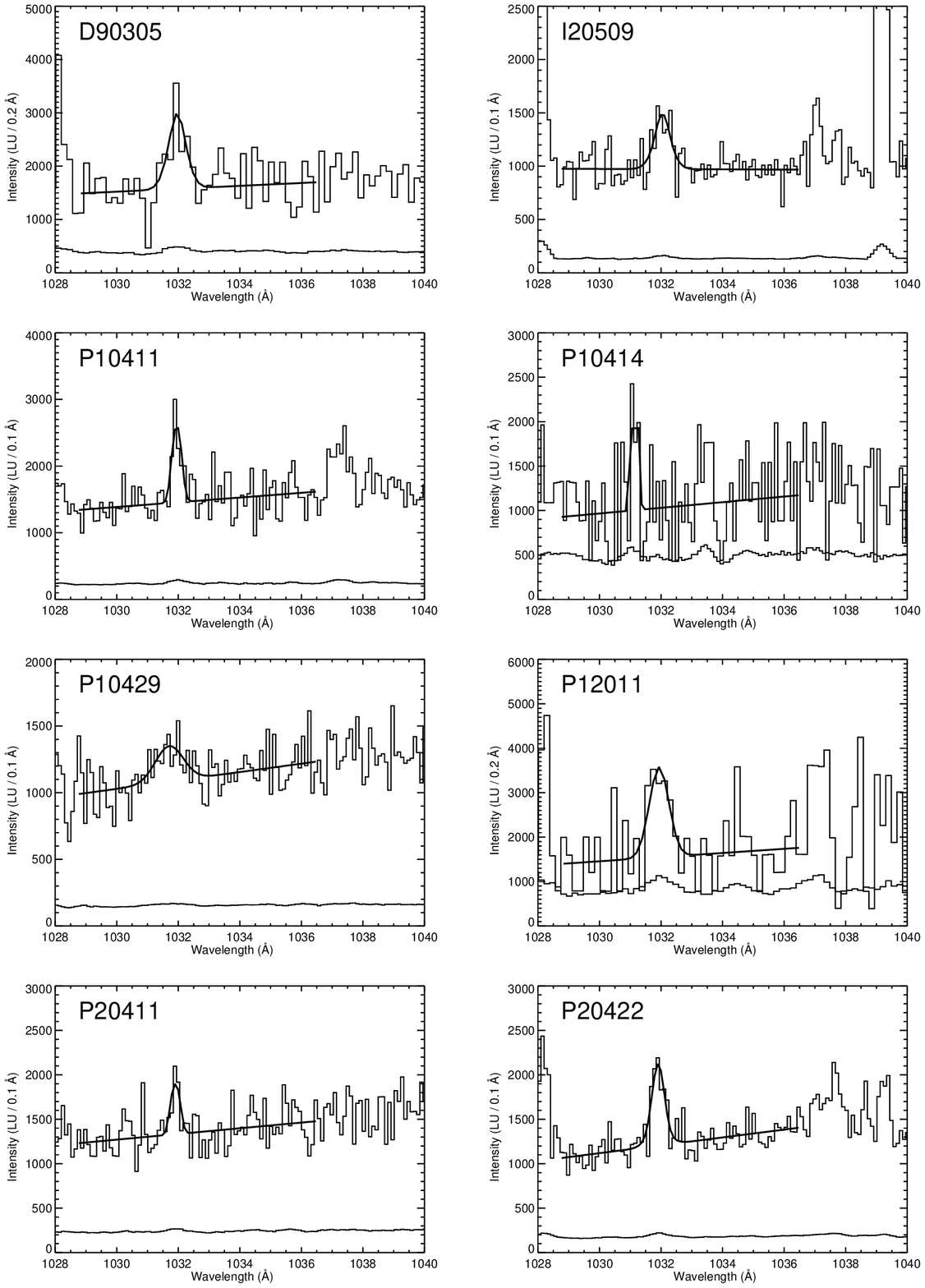}
\centerline{{\it Fig. 2. --- Continued.}}
\clearpage
\plotone{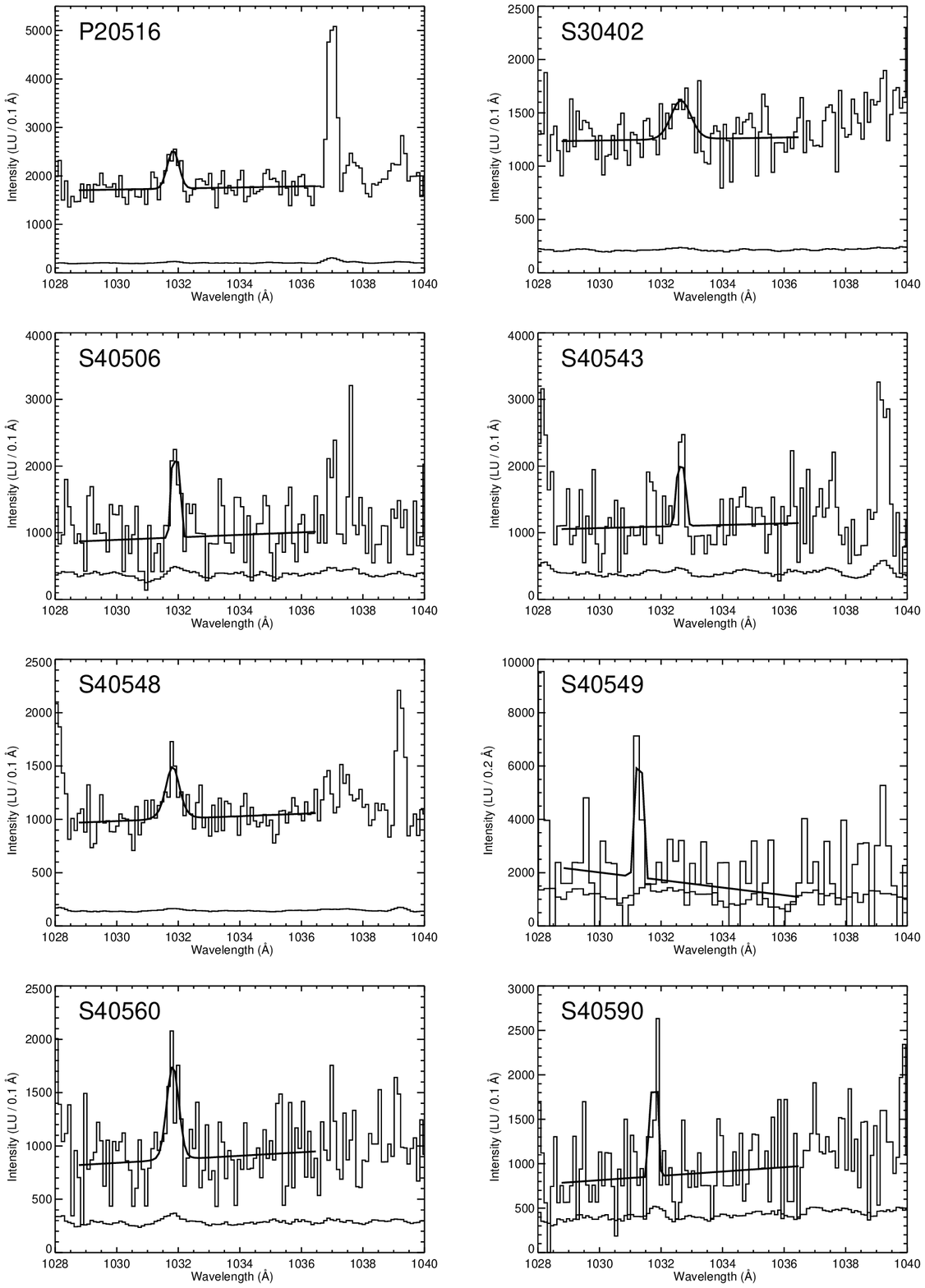}
\centerline{{\it Fig. 2. --- Continued.}}
\clearpage
\plotone{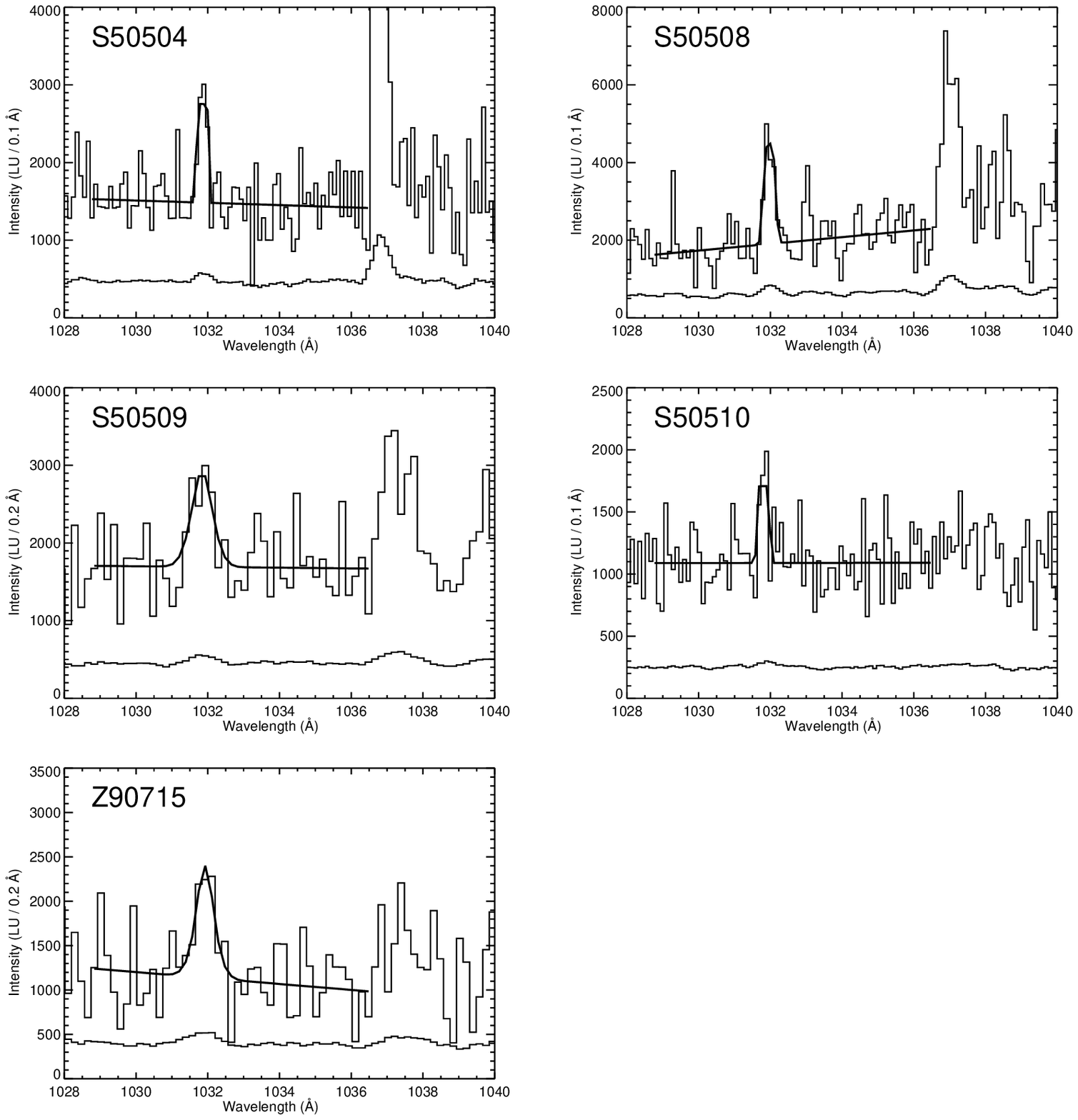}
\centerline{{\it Fig. 2. --- Continued.}}
\clearpage
\begin{figure}
\figurenum{3}
\plotone{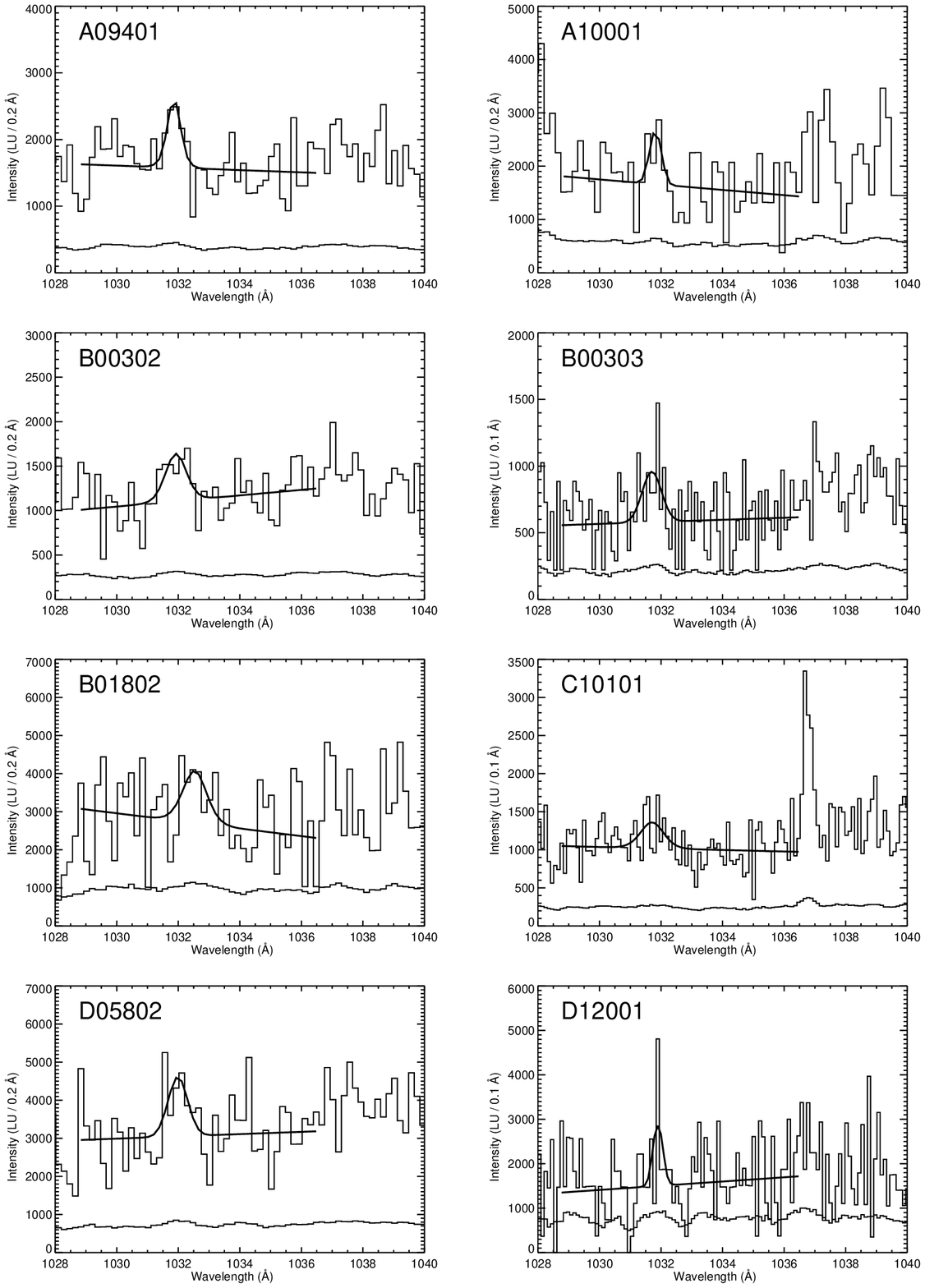}
\caption{Low-significance (2\sig) \osix\ $\lambda 1032$ emission features.
The data are binned by the factors given in Table \ref{tab_detections}, and 
best-fit model spectra and error bars are overplotted.
Interstellar \ctwo * $\lambda 1037$ and \osix\ $\lambda 1038$ are present in some spectra,
as are the geocoronal \oone\ $\lambda \lambda 1028, 1039$ lines.}
\label{fig_2sigma}
\end{figure}
\clearpage
\plotone{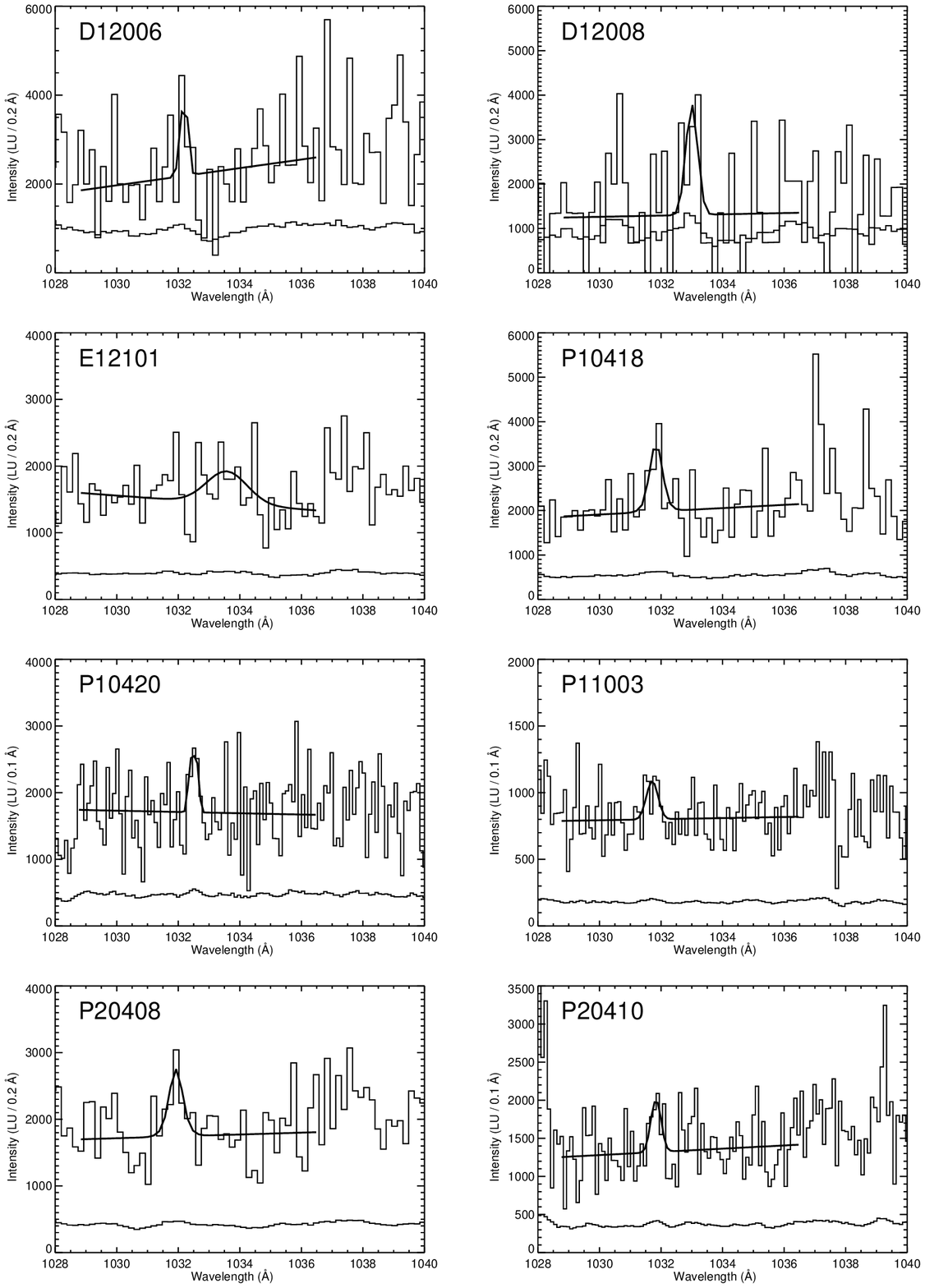}
\centerline{{\it Fig. 3. --- Continued.}}
\clearpage
\plotone{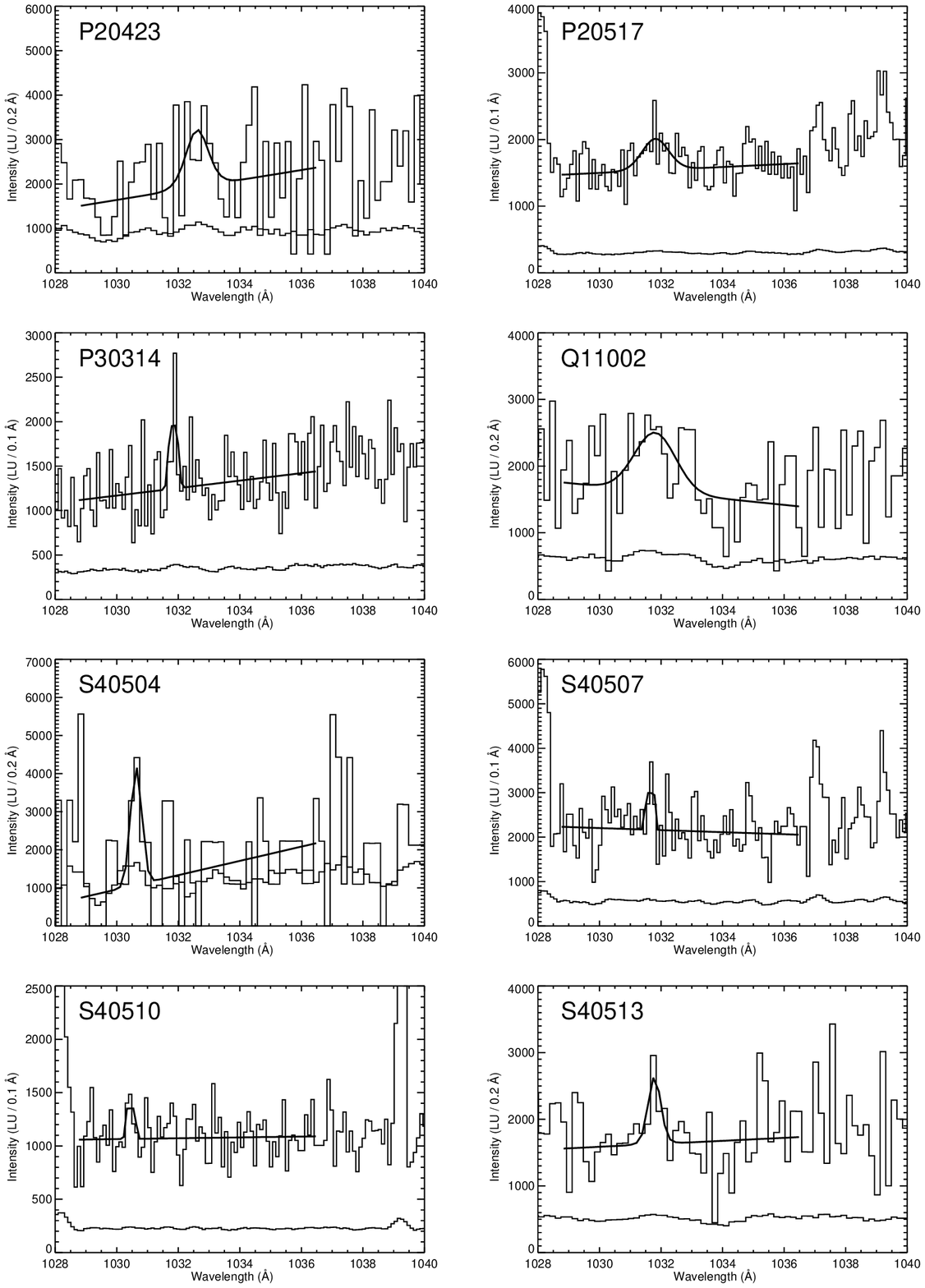}
\centerline{{\it Fig. 3. --- Continued.}}
\clearpage
\plotone{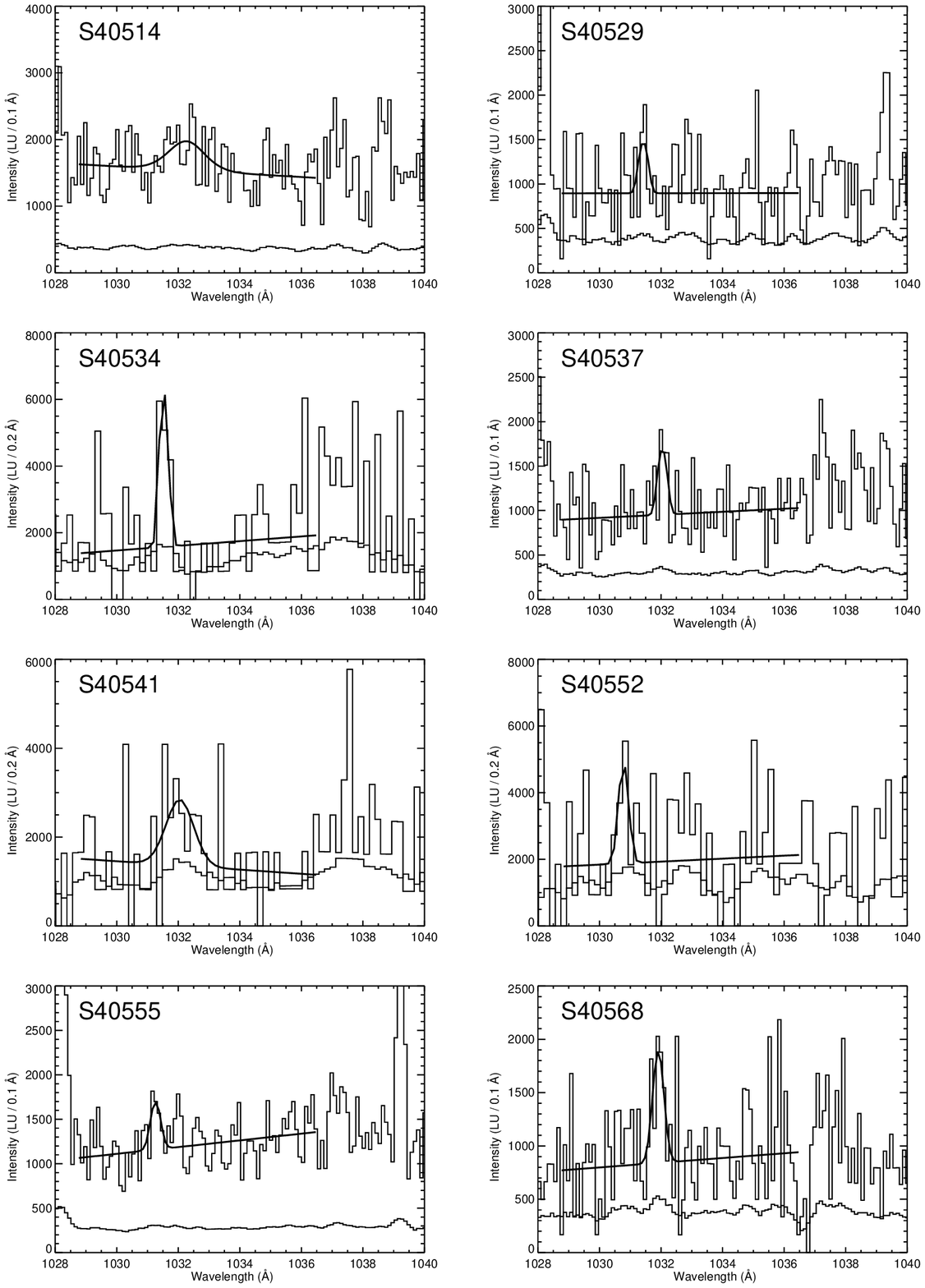}
\centerline{{\it Fig. 3. --- Continued.}}
\clearpage
\plotone{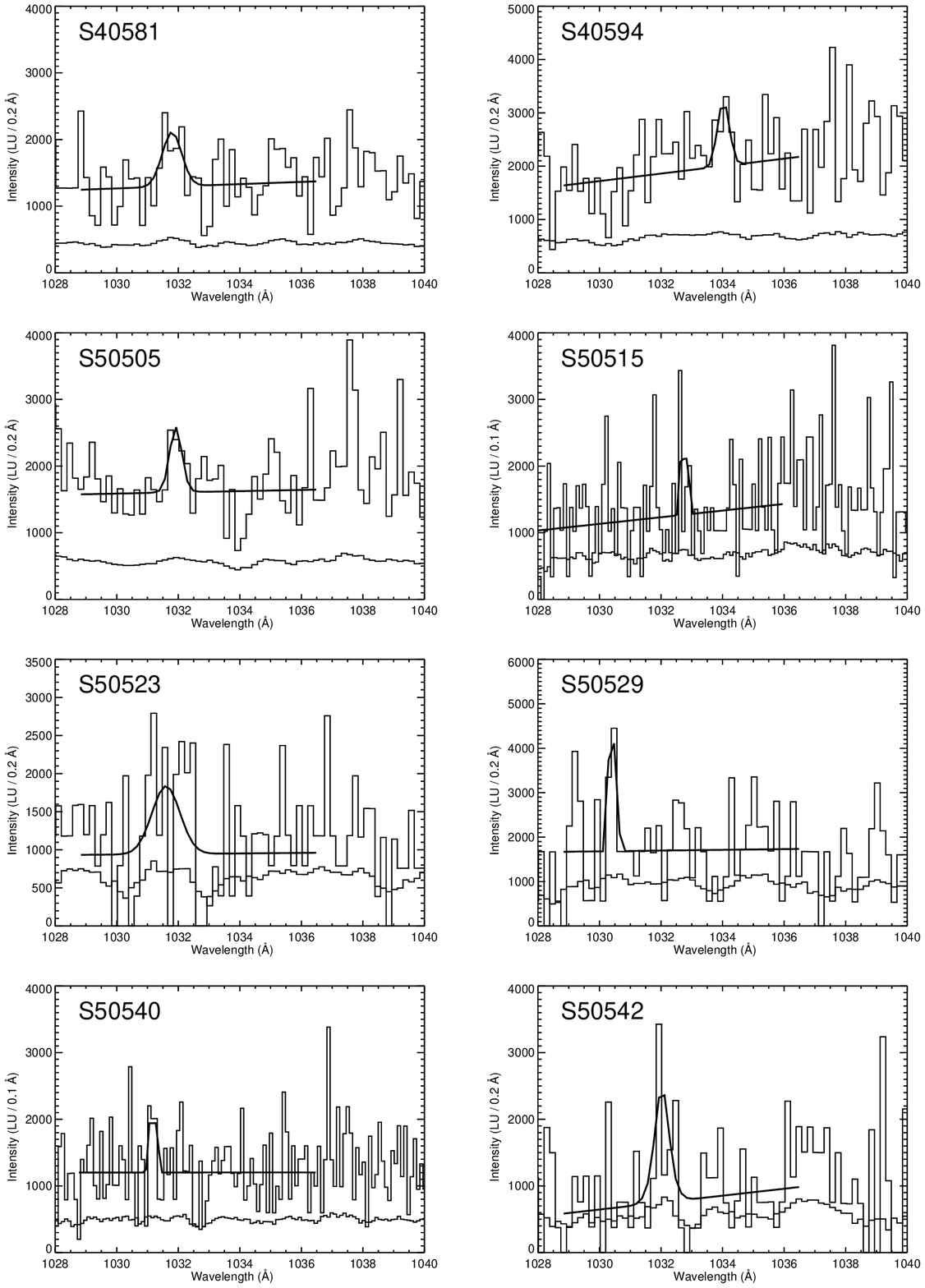}
\centerline{{\it Fig. 3. --- Continued.}}
\clearpage
\plotone{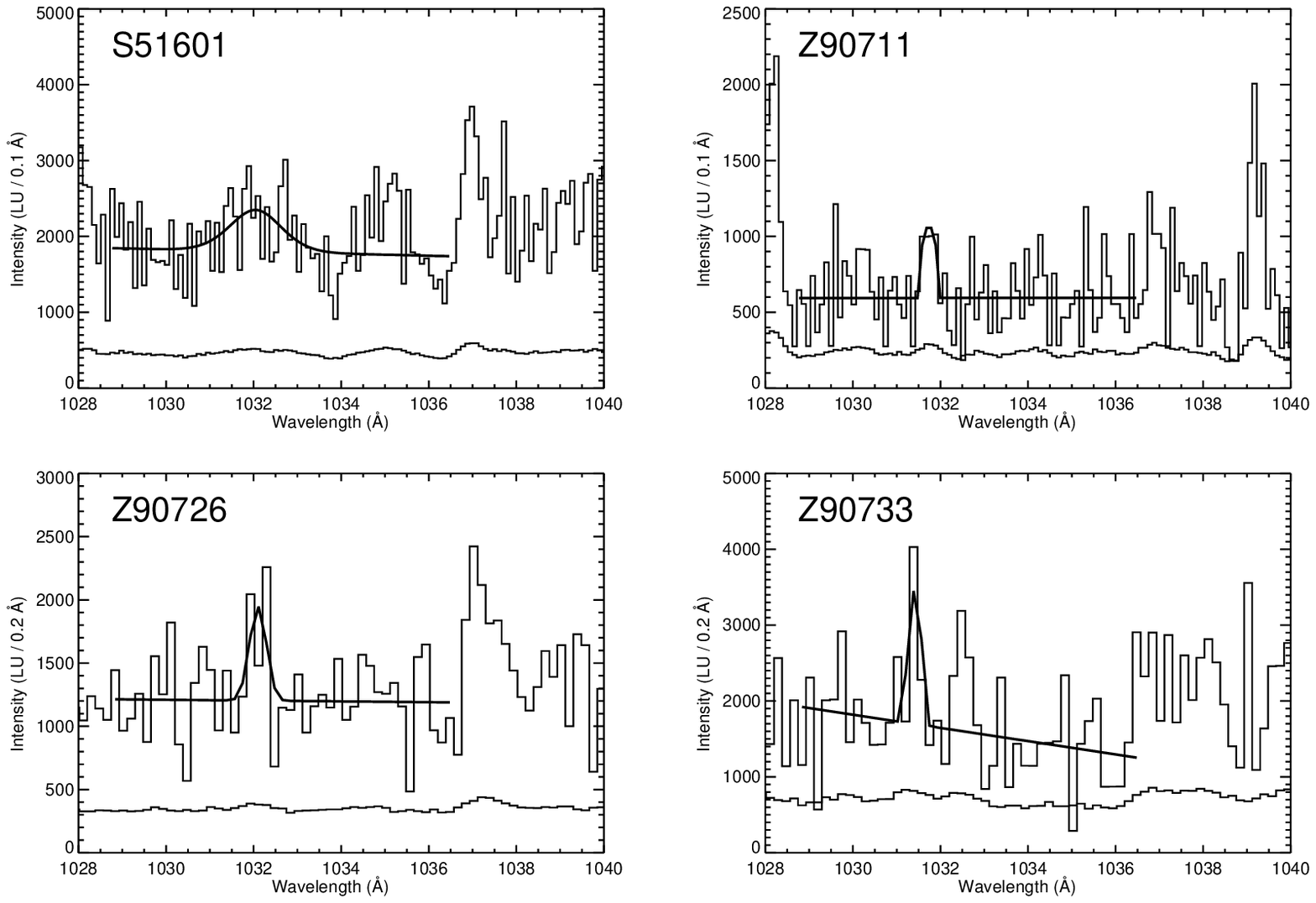}
\centerline{{\it Fig. 3. --- Continued.}}

\end{document}